\documentclass[a4paper,onecolumn,superscriptaddress,nofootinbib,accepted=2021-05-11]{quantumarticle}
\pdfoutput=1
\usepackage[latin1]{inputenc}
\usepackage[english]{babel}
\usepackage{hyperref}
\usepackage{amsmath,color}
\usepackage{amsfonts}
\usepackage{amssymb,braket,tikz-cd,amsmath,linegoal}
\usepackage{graphicx}
\usepackage{mdframed}
\usepackage{amsthm}
\usepackage{tabularx}
\usepackage{mathtools}
\usepackage{geometry}
\usepackage[mathscr]{euscript}
\usepackage{theoremref}
\usepackage{changes}
\usepackage[inline]{enumitem}
\usepackage[numbers,sort&compress]{natbib}

\theoremstyle{definition}
\newtheorem{definition}{Definition}
\newtheorem{lemma}{Lemma}
\newtheorem{theorem}{Theorem}

\newtheorem{example}{Example}

\newtheorem{corollary}{Corollary}
\newtheorem{remark}{Remark}
	
\def\u{\mathbf u}
\def\f{\mathbf f}

\def\fF{\mathfrak F}

\def\F{\mathcal F}
\def\M{\mathcal M}
\def\D{\mathcal D}
\def\S{\mathcal S}

\def\C{\mathbb C}
\def\R{\mathbb R}

\def\I{\mathbb I}
\def\H{\mathbb H}
\def\bF{\mathbb F}
\def\Z{\mathbb Z}

\def\0{\bf 0}

\def\P{{\rm P}}
\def\PC{{\rm P} \mathbb{C}}
\def\PR{{\rm P} \mathbb{R}}
\def\PH{{\rm P} \mathbb{H}}
\def\SU{{\rm SU}}
\def\PU{{\rm PU}}
\def\U{{\rm U}}
\def\SO{{\rm SO}}
\def\O{{\rm O}}
\def\bO{ \mathbb{O}}
\def\Sp{{\rm Sp}}
\def\GL{{\rm GL}}
\def\Hom{{\rm Hom}}
\def\Ind{{\rm Ind}}
\def\dim{{\rm dim}}
\def\Diff{{\rm Diff}}

\def\Sym{{\rm Sym}}
\def\sp{{\rm span}}
\def\conv{{\rm conv}}
\def\spann{{\rm span}}

\def\Gr{{\rm Gr}}
\def\dist{{\rm dist}}
\def\sym{{\rm sym}}

\def\tr{{\rm tr}}

\def\Quart{\tt Quart}

\def\Quant{{\tt Quant}}

\definechangesauthor[name=Thomas, color=red]{TG}
\definechangesauthor[name=Lluis, color=blue]{LM}

\begin{document}

\title{How dynamics constrains probabilities in general probabilistic theories}
\author{Thomas D. Galley}	
\email{tgalley1@perimeterinstitute.ca}
\affiliation{Perimeter Institute for Theoretical Physics, Waterloo, ON N2L 2Y5, Canada}
\author{Llu\'\i s Masanes}
\affiliation{Department of Computer Science, University College London, Gower Street, London WC1E 6BT, United Kingdom}

\begin{abstract}
We introduce a general framework for analysing general probabilistic theories, which emphasises the distinction between the dynamical and probabilistic structures of a system. The dynamical structure is the set of pure states together with the action of the reversible dynamics, whilst the probabilistic structure determines the measurements and the outcome probabilities. For transitive dynamical structures whose dynamical group and stabiliser subgroup form a Gelfand pair we show that all probabilistic structures are rigid (cannot be infinitesimally deformed) and are in one-to-one correspondence with the spherical representations of the dynamical group. 
We apply our methods to classify all probabilistic structures when the dynamical structure is that of complex Grassmann manifolds acted on by the unitary group.
This is a generalisation of quantum theory where the pure states, instead of being represented by one-dimensional subspaces of a complex vector space, are represented by subspaces of a fixed dimension larger than one. We also show that systems with compact two-point homogeneous dynamical structures (i.e.~every pair of pure states with a given distance can be reversibly transformed to any other pair of pure states with the same distance), which include systems corresponding to Euclidean Jordan Algebras, all have rigid probabilistic structures.  
\end{abstract}

\maketitle

\section{Introduction}

\emph{General probabilistic theories} (GPTs) provide a framework for the study of operational theories beyond quantum theory. Within this framework quantum theory appears as one non-classical theory amongst many. This field has its origin in the work of Segal~\cite{Segal_postulates_1947}, Mackey~\cite{Mackey_mathematical_1963} and Ludwig~\cite{Ludwig_versuch_1964,Ludwig_attempt_1967,Ludwig_attempt_1968} with other notable contributions  at the time including~\cite{dahn_attempt_1968,stolz_attempt_1969,stolz_attempt_1971,Mielnik_theory_1969,Mielnik_generalized_1974,giles_foundations_1970,
Gudder_convex_1973,Davies_operational_1970} amongst others.  Contemporary interest in GPTs was kickstarted by Hardy's seminal work~\cite{Hardy_quantum_2001} followed by a detailed exposition of the framework by Barrett~\cite{Barrett_information_2005}. Important applications of the framework include the operational derivations of quantum theory of~\cite{Dakic_quantum_2011,Masanes_derivation_2011,Chiribella_informational_2011}. Current treatments have tended to emphasise finite dimensional systems and system composition. Using this framework (or related frameworks such as convex operational theories~\cite{barnum_information_2011, barnum_symmetry_2010} and operational probabilistic theories~\cite{Chiribella_probabilistic_2010}) many physical and informational features of general probabilistic theories have been studied, such as interference phenomena~\cite{garner_framework_2013,barnum_higher_order_2014, barnum_ruling_2017,Lee_higher_2017}, computation~\cite{Muller_power_2011,lee_computation_2015}, thermodynamics~\cite{Brunner_dimension_2014,Chiribella_operational_2015,chiribella_entanglement_2015} and others~\cite{Barnum_entropy_2010,kimura_distinguishability_2010,Janotta_limits_2011,Janotta_gen_2013,Barnum_composites_2020,bae_structure_2016,selby_how_2018,heinosaari_no_2019}. 

Examples of GPTs (excepting  classical and quantum theory) include Boxworld~\cite{davies_example_1972,Barrett_information_2005,Short_strong_2010,Gross_all_2010,Sabri_reversible_2014}, quantum theory over the field of real numbers~\cite{Hardy_limited_2012,Aleksandrova_real_2013,Wootters_optimal_2013} or quaternions~\cite{finkelstein_notes_1959}, theories based on Euclidean Jordan algebras~\cite{Barnum_composites_2020}, quartic quantum theory~\cite{Zyczkowski_quartic_2008}, $d$-balls~\cite{Dakic_quantum_2011,Masanes_entanglement_2014,krumm_quantum_2019}, density cubes~\cite{Dacik_density_2014} and quantum systems with modified measurements~\cite{Galley_classification_2017}. Amongst these, only Boxworld, quantum theory over real or quaternionic fields and theories based on Euclidean Jordan algebras are full theories, in that they have non-trivial composites. 

The aim of this paper is to provide tools to systematically explore the space of non-classical systems. Rather than generating examples of non-classical systems we can give full classifications of families of non-classical systems which share a common dynamical structure (pure states and reversible dynamics) but different probabilistic structures (measurements and measurement outcome probabilities); as done in~\cite{Galley_classification_2017} for systems which share the dynamical structure of quantum systems. We can thus obtain a richer picture of the space of non-classical systems, of which quantum systems are just one example.

We provide a general framework for convex systems and use it to study \emph{transitive} systems, that is to say systems for which any two pure states are related by a reversible transformation. This is a generalisation of the OPF (outcome probability function) framework of~\cite{Galley_classification_2017,galley_modification_2018, masanes_measurement_2019}, where the pure states and dynamical group no longer have to be those of quantum theory. We restrict ourselves to systems with reversible dynamics given by finite and compact groups, noting that all the examples of GPT systems mentioned previously are transitive systems with finite or compact dynamical groups. The assumption of transitivity has played an important role in derivations of quantum theory from operational/information theoretic principles, for instance in Hardy's original derivation~\cite{Hardy_quantum_2001} as well as subsequent derivations by other authors~\cite{wilce2009half,Dakic_quantum_2011,Masanes_derivation_2011,Chiribella_informational_2011}.
It is worth mentioning that many derivations of the second law of thermodynamics from more fundamental principles (see for example~\cite{Rio_2011,Horodecki_2013,Brand_o_2013,Skrzypczyk_2014,Brand_o_2015}) use as the central premise the reversibility of the underlying dynamics (both in the classical and quantum frameworks).
Also, when all the transformations that can be implemented on a system are generated by reversible dynamics, all the achievable states of the system form a transitive space. Hence, there is a connection between transitivity and the second law of thermodynamics.

We show that for a given dynamical structure (pure states and dynamical group) every possible probabilistic structure (measurements and outcome probabilities) is in correspondence with a representation of the dynamical group. Moreover we find necessary and sufficient conditions on the dynamical structure (the dynamical group and subgroup form a Gelfand pair) which make this correspondence one-to-one. We find that certain probabilistic structures cannot be infinitesimally deformed and call these rigid. We show that all dynamical structures which are Gelfand pairs do not have any probabilistic structures which can be infinitesimally deformed. We apply the methods developed to classify generalisations of quantum systems, with pure states given by Grassmann manifolds and unitary dynamics. We introduce the family of systems with compact two point homogeneous dynamical structures and show that they all have rigid probabilistic structures.

\subsection{Structure of the paper}

In Section~\ref{sec:single_sys} we introduce the OPF framework used for studying transitive systems and present relevant known results (or slight generalisations thereof). 
In Section~\ref{sec:classification} we give the main theorem of this work (the classification theorem), establishing a correspondence between probabilistic structures of transitive systems and group representations, as well as the conditions under which this correspondence is one-to-one. 
In Section~\ref{sec:rigid} we introduce the notion of deformation of probabilistic structures, and show that the only dynamical structures which admit probabilistic structures which can be infinitesimally deformed are those corresponding to non-Gelfand pairs. We also give an explicit example of deformations of a non-rigid probabilistic structure.
In Section~\ref{sec:twopoint} we introduce the family of compact dynamical structures which are two point homogeneous and show that they are all rigid.
In Section~\ref{sec:grassmann} we apply the classification theorem to systems with dynamical structures given by complex Grassmann manifolds (a generalisation of complex projective space).
In Section~\ref{sec:discussion} we discuss the results of this paper in light of existing work as well as comment on the implications of new concepts and results of the present work. 
Lastly we close with some concluding remarks in Section~\ref{sec:conclusion}.
A glossary of notation is given in Appendix~\ref{app:notation} and an introduction to some of the representation theory used in this paper can be found in Appendices~\ref{app:group_rep} and \ref{app:real_rep}.

\section{Single system state spaces}\label{sec:single_sys}

We provide a characterisation of single systems within the GPT framework which emphasises the pure states and reversible dynamics. This will allow us to consider families of systems with the same pure states and reversible dynamics, but different measurements. This is a generalisation of~\cite{Galley_classification_2017} where all systems with the same pure states and reversible dynamics as quantum theory were classified and their informational properties studied. We first describe quantum systems in this framework.

\subsection{A characterisation of finite dimensional quantum systems}

Quantum systems are often characterised directly in terms of mixed states, that is to say their convex representation. States are positive semi-definite operators on a finite dimensional complex space $\C^d$ equipped with a sesquilinear inner product, transformations are CPTP maps and measurements are associated to POVMs, with the probability of an outcome occuring being given by the usual trace rule. Here we provide a characterisation of quantum systems which separates their dynamical structure from the probabilistic structure. In this characterisation the mixed state representation described above is derived, rather than postulated. Moreover this distinction between dynamical and probabilistic structures will provide us with a way of classifying families of more general systems which share a common dynamical structure. A quantum system $\S^\Quant_d$ associated to the space $\C^d$ is given by the following:

\begin{center}
\begin{enumerate}[label = \Roman*. , align = left]
\item Pure states: $\psi \in \PC^d$.
\item Reversible dynamics:  $\psi \mapsto U \psi$, $U \in \PU (d)$.
\item Outcome probabilities: $\F^\Quant_{\PC^d}=\{Q:\PC^d \to [0,1]\, |\, Q(\psi) = \braket{\psi |\hat Q|\psi}, \, \forall\, \hat Q: 0\leq \hat Q \leq \I \}$.
\end{enumerate}
\end{center}

$\PC^d$ is the complex projective space with elements corresponding to one dimensional subspaces of $\C^d$. $\PU(d)$ is the projective unitary group constructed from $\U(d)$ by taking equivalence classes of unitaries under multiplication by a complex phase: $U_0 = e^{i \theta} U_1 \iff U_0 \cong U_1$ where $U_0, U_1 \in \U(d)$ and $\theta \in \R$.

We assume that any subset $\{Q_i\}_{i = 1}^n$ such that $\sum_i Q_i(\psi) = 1$ for all $\psi\in \PC^d$ forms a valid measurement. This implies that a measurement consists of positive semi-definite operators $\hat Q_i$ such that $\sum_i \hat Q_i = 1$. Here I. and II. are the dynamical structure, whilst III. is the probabilistic structure. The mixed state representation (density operators) is derived from the dynamical and probabilistic structures. We observe that the probability assignment (Born rule) is not given in terms of the trace, since this already presumes the structure of mixed states. We now define general non-classical systems in terms of dynamical and probabilistic structures and show how to derive the convex representation.

\subsection{Dynamical structure}

The pure states of a system $\S$ form a set $X$, and the reversible dynamics a group $G$. The action of $G$ on $X$ is given by a group action $\varphi: G \times X \to X$. This gives $X$ the structure of a $G$-space. 
\begin{definition}[Dynamical structure]
A dynamical structure $\mathcal D$ is a triplet
\begin{equation}
\mathcal{D} =(X , G ,  \varphi),
\end{equation}
where $X$ is a set, $G$ is a group and $\varphi$ a group action. 
\end{definition}
In the following we leave $\varphi$ implicit and write $gx$ for $\varphi(g,x)$.
An important family of dynamical structures are transitive. A dynamical structure is \emph{transitive} when for any two pure states $x,x' \in X$ there exists a transformation $g \in G$ such that $x' = gx$.
In other words $X$ is the orbit of $G$ acting on an arbitrary $x \in X$.
A central notion to the approach used in this work is that of a \emph{stabilizer subgroup} (also known as isotropy group) of an element $x \in X$, which is just the subgroup of all transformations in $G$ which leave a point $x$ invariant. We write $H_x := \{g \in G : gx = x \}$ for the stabilizer subgroup of a point $x \in X$.  For a transitive group action, the stabilizer groups for different points are isomorphic, hence we write $H$ as the stabilizer group.

Given a group $G$ and a subgroup $H\subseteq G$, we denote by $\phi_{G,H}$ the action of $G$ on the set of left cosets $G/H$. 
Given a transitive dynamical structure $\mathcal D = ( X, G, \varphi)$ with stabilizer subgroup $H$ we have the following isomorphism of dynamical structures $( X, G, \varphi) \cong (G/H, G, \phi_{G,H})$. 

Typically dynamical structures $( X, G, \varphi)$ also have topological (and sometimes differentiable) structure. A topological group $G$ acts on a topological space $X$ when the action $\varphi$ is a continuous function: $\varphi: G \times X \to X$ (where $G \times X$ has the product topology). If $H$ is a subgroup of $G$ then the space $G/H$ of left cosets is a topological space with respect to the quotient topology, which is the finest topology making the quotient map $q: G \to G/H$, $q: g \mapsto gH$ continuous. 

For transitive dynamical structures $( X, G, \varphi)$ where $X$ is Hausdorff and $G$ is compact the isomorphism $( X, G, \varphi) \cong (G/H, G, \phi_{G,H})$ also involves the topological structure of each component of the triplet~\cite[Proposition 1.10]{Hofman_structure_2013}. When $G$ is a Lie group the isomorphism $( X, G, \varphi) \cong (G/H, G, \phi_{G,H})$ involves the differentiable structure of each component of the triplet~\cite[Theorem 20.12]{gallier_advanced_2018}.

In the following we restrict our attention to compact dynamical structures (which includes finite dynamical structures as a special case) implying that the isomorphism $(X, G, \varphi) \cong (G/H, G, \phi_{G,H})$ includes the topological and differentiable structures of the dynamical structures considered. 
 
For this reason we use the abbreviation
\begin{equation}
\mathcal{D} = (X,G,\varphi) =(G,H)\ .
\end{equation}
for transitive dynamical structures with stabilizer subgroup $H$.

The case of non-compact dynamical structures is discussed in Section~\ref{sec:non-comp_dyn_struc}.

\subsection{Probabilistic structure}

A system is determined by its pure states, dynamics and measurements. Given the dynamical structure we need to specify its probabilistic structure, which characterises the measurements which can be performed on the system. 

\begin{definition}[Outcome probability function (OPF)]
An outcome of a measurement on a system with pure states $X$ is given by a function $\f: X \to [0,1]$, where the probability of the associated outcome $\f$ occurring is $P(\f|x) = \f(x)$. 
\end{definition}

\begin{definition}[Measurement]
A measurement  $\M_j$ with  countable outcomes $i = 1, ... , n,..$ is specified by the list $\{ \f_1^j, ..., \f_n^j,... \}$. The elements of this list obey the condition:
\begin{equation}
\sum_i \f^j_i(x) = 1, \ \forall x \in X \ .
\end{equation}
\end{definition}

\begin{definition}[Unit OPF]
The unit OPF $\u$ is $\u(x) = 1 , \forall x \in X$. 
\end{definition}

\begin{definition}[Probabilistic structure]
The probabilistic structure of a system is the set $\F_X$ of all outcome probability functions $\f$.
\end{definition}
Typically we assume that any set $\{\f_1, ..., \f_n,...\}$ such that $\sum_i \f_i = \u$ forms a valid measurement, however this assumption is not necessary. When this assumption does not hold, one needs to supplement the set $\F_X$ with a specification of which OPFs form a valid measurement. One example of such a specification is the `finite measurement outcomes' assumption:
\begin{definition}[Finite measurement outcome assumption]
	Only finite sets of OPFs $\{\f_1, ..., \f_n\}$ such that $\sum_i \f_i = \u$ form valid measurements.
\end{definition}
The above assumption is sometimes viewed as part of the definition of measurements in an operational framework, since we can never carry out measurements with infinitely many outcomes. We will be making this assumption in the present work. 

Operational considerations impose the following constraints on $\F_X$:
\begin{enumerate}[label = \roman* . ,align = left]
\item $\F_X$ is closed under taking mixtures: for all $\f_1, \f_2 \in \F_X$ and all $\lambda \in [0,1]$ we have that $\lambda \f_1 + (1-\lambda) \f_2 \in \F_X$. 

\item $\F_X$ is closed under composition with group transformations: for all $\f \in \F_X$ and $g \in G$ we have that $\f \circ g \in \F_X$, where $(\f\circ g)(x) = \f(gx)$.

\item $\F_X$ is closed under coarse graining of measurement outcomes: for any pair of outcomes $\f_i^k, \f_j^k \in \M_k$ of a given measurement $\M_k$ we have that $\f_i^k + \f_j^k \in \F_X$.
\item For every $\f \in \F_X$, the complement OPF $\f_c = \u - \f$ is also in $\F_X$
\end{enumerate}
The first constraint implies that $\F_X$ is a convex set, hence it can be extended to a vector space $\R[\F_X]$ with addition $(\f_1 + \f_2)(x) = \f_1(x) + \f_2(x)$ and scalar multiplication $(\alpha \f)(x) = \alpha \f(x)$ for any $\alpha \in \R$. Explicitly $\R[\F_X]: = \{\tilde \f(x)| \tilde \f(x) = \alpha \f_1(x) + \beta \f_2(x), \ \alpha, \beta \in \R, \ \f_1, \f_2 \in \F$. We have that $\F$ is a convex subset of $\R[\F_X]$ by construction and that $\spann_\R(\F_X) = \R[\F_X]$.

Closedness under composition with group transformations implies that $\F_X$ is a $G$-space.
This and the fact that the group action commutes with taking mixtures implies that $\R[\F_X]$ is a linear representation of $G$.
Closedness under coarse graining of measurement outcomes implies that every $\F_X$ contains the unit OPF and the existence of the complement guarantees the existence of the $\bf 0$ OPF.
We introduce the following property, though we will not always require it in the present treatment.
\begin{definition}[Separability of pure states]\label{PureSep}
A probabilistic structure $\F_X$ separates pure states when  for any two pure states  $x_1 , x_2 \in X$ there exists an OPF $\f \in \F_X$ such that $\f(x_1) \neq  \f(x_2)$.
\end{definition}
If one does not have this requirement, the probabilistic structure $\F_X = \{\u\}$ leading to a trivial system for all dynamical structures is valid for example.

\subsection{Systems, state spaces and associated group representations}

The above definitions allow us to formally define a system $\S_X$.

\begin{definition}[System]
	A system $\S_X$ is a triple $\S_X = (X,G , \F_X)$, where $(X,G)$ is a dynamical structure and $\F_X$ is a probabilistic structure.
\end{definition}

In the following we briefly outline how the general state space (including mixed states) of a system is derived, both from an operational starting point and directly from the mathematical starting point $\S_X = \{X,G , \F_X\}$.  

\subsubsection{Operational derivation of the state space}\label{subsubsec:op_deriv}

Operationally for a single system one has access to a preparation device  which is wired up sequentially with a transformation and measurement devices. These devices have classical settings (for instance which transformation to apply) and classical readouts (for instance which measurement outcome occurred). In an experiment one collects the statistics for different outcomes given choices of settings. Typically one assumes that statistics are gathered for all possible setting choices, and that the relative frequencies obtained become probabilities as the number of runs tends to infinity. Using these probabilities (which are directly given by the set $\F$ in the OPF framework) one derives the convex state space (and effect space) of the system. We refer the reader to~\cite{mazurek_experimentally_2017} about how one can in practice derive a state and effect space from experimental data.

\subsubsection{Mathematical derivation of the state space}\label{subsubsec:mat_deriv}

In this work we will make the assumption of the possibility of state estimation using a finite outcome set (known as `Possibility of state estimation' in~\cite{masanes_measurement_2019}).
\begin{definition}[Possibility of state estimation using a finite outcome set]
	The system $\S = \{X,G,\F_X\}$ is such that the value of a finite number of outcomes $\f_1, ..., \f_n \in \F_X$ on any ensemble $\{(p_i , x_i)\}_i$ determines the value of any OPF $\f \in \F_X$ on the ensemble $\{(p_i , x_i)\}_i$.
\end{definition}
It is shown in Lemma 2 of~\cite{masanes_measurement_2019} that this implies that $\mathbb R[\F_X]$ is finite dimensional. Equivalently the convex set of mixed stated is embeddable in a finite dimensional real vector space.
We now briefly outline the derivation of the space of mixed state for a system  $\S_X = \{X,G,\F_X\}$ under the assumption  ``Possibility of state estimation using a finite outcome set''.
First the probability of an outcome $\f$ (defined on $X$) occurring for an ensemble $\{(p_i , x_i)\}_i$ is $P(\f|\{(p_i , x_i)\}_i) = \sum_i p_i \f(x_i)$. This allows us to define \emph{equivalent} ensembles.
\begin{definition}[Equivalent ensembles]
Two ensembles  $ \{(p_i , x_i)\}_i$ and $ \{(p_j' , x_j')\}_j$ are equivalent if $P(\f|\{(p_i , x_i)\}_i) =  P(\f|\{(p_j' , x_j')\}_j)$ $\forall  \f \in  \F_X$, and we write $\{(p_i , x_i)\}_i \sim \{(p_j' , x_j')\}_j$.
\end{definition}
The mixed states are defined as equivalence classes of ensembles under this equivalence relation. 
For each state $x\in X$ we define the linear functional $\Omega_x : \mathbb R [\F_X] \to \mathbb R$ as $\Omega_x (\f) = \f(x)$. The probability of outcome $\f$ on ensemble $\{(p_i , x_i)\}_i$ can be written as
\begin{equation}
  P(\f|\{(p_i , x_i)\}_i)
  =
  \sum_i p_i \f(x_i)
  =
  \sum_i p_i \Omega_x(\f)
  =
  \omega(\f)\ ,
\end{equation}
where we define the functional associated to ensemble $\{(p_i , x_i)\}_i$ as $\omega = \sum_i p_i \Omega_x$.
Therefore, two ensembles$ \{(p_i , x_i)\}_i$ and $ \{(p_j' , x_j')\}_j$  are equivalent if and only if, their corresponding functionals are identical
$\sum_i p_i  \Omega_{x_i} = \sum_j p_j'  \Omega_{x_j'}$. The outcome probabilities $P(\f|\{(p_i , x_i)\}_i)$ on the space of ensembles uniquely define linear functionals $\Lambda_\f$ on the space of mixed states, such that $\Lambda_\f \cdot \omega = \omega(\f)$ for all mixed states $\omega$.

The group action $\varphi: X \times  G \to X$, naturally extends to the space of mixed states (embedded linearly in $\R[\F_X]^*$) as $\Omega_x \xrightarrow{g} \Omega_{gx}$. This is a linear action which is such that $\Omega_x \xrightarrow{g} \Omega_{gx} \xrightarrow{g'} \Omega_{g'gx}$ is the same as $\Omega_x  \xrightarrow{g'g} \Omega_{g'gx}$; hence there exists a homomorphism $\Gamma: G \mapsto \GL(\R[\F_X]^*)$. We call this the group representation \emph{associated} to the system $\S$.  This naturally induces a representation $\Gamma^*: G \mapsto \GL(\R[\F_X])$, which is isomorphic to $\Gamma$ since the representations are unitary and real.

We can summarise the above in the following theorem (fully proven in Appendix~\ref{app:statespace}), which is a straightforward generalisation of Result 1 of~\cite{Galley_classification_2017} to arbitrary dynamical structures:

\begin{theorem}[Result 1 of~\cite{Galley_classification_2017}]\label{thm:sys_maps}
	For every system $\S_X = \{X, G , \F_X\}$ obeying `Possibility of state estimation using a finite outcome set' there exists an embedding of $\S_X$ into a finite dimensional real vector space $V \cong \R[\F_X]^*$ and its dual $V^*$ given by the following maps:
	\begin{align}
	\Omega &: X \to V \\
	\Gamma &: G \to \GL(V) \label{eq:rep_eq}  \\
	\Lambda &: \F_X \to V^* 
	\end{align}
	satisfying the following properties:
	\begin{enumerate}
		\item Preservation of dynamical structure:
		\begin{align}
		\Gamma_g \Omega_x & = \Omega_{gx} \\
		\Gamma_{g_1} \Gamma_{g_2} &  = \Gamma_{g_1 g_2}
 		\end{align}
 		\item Preservation of probabilistic structure:
 		\begin{align}
 		\Lambda_\f \cdot \Omega_x = \f (x)
 		\end{align}
 		\item Uniqueness: The embedding of $\S_X$ into $(V,V^*)$ given by the maps $\Omega, \Gamma,\Lambda$ (satisfying all of the above) is unique up to equivalence. 
 		
 		Two embeddings of $\S_X$ into $(V,V^*)$ given by the maps  $\Omega, \Gamma,\Lambda$ and $\Omega', \Gamma',\Lambda'$  are equivalent if there exists an invertible linear map $L: V \to V$ such that:
 		\begin{align}
 		\Omega_x' & = L \Omega_x, \ \forall x \in X , \\
 		\Gamma_g ' & = L \Gamma_g L^{-1}, \ \forall g \in G, \\
 		\Lambda_\f'& = \Lambda_\f L^{-1}, \forall \f \in \F_X \ .
 		\end{align}
	\end{enumerate}
\end{theorem}

We call the representation $\Gamma$ of Equation~\eqref{eq:rep_eq} the representation of $G$ \emph{associated} to the system $\S_X$. $\conv(\Omega_X)$ is the convex hull of the extremal states, which we call \emph{state space}~\footnote{The object $\conv(\Omega_{X}) = \conv(\Gamma_G \Omega_{x})$ is the convex hull of an orbit of the group $G$ in the vector space $V$ and is known as an \emph{orbitope}, see~\cite{Sanyal_2011} for a detailed study of the theory of orbitopes.}. A standard representation of the states $\Omega_X$ is given as a vector of fiducial outcome probabilities. 

\begin{remark}
For a system $\S_X = \{X, G , \F_X\}$ with maps $\Omega, \Gamma,\Lambda$ the vectors $\Omega_X$ admit of a standard representation in terms of a fiducial outcome set (which is non-unique). A fiducial outcome set $\{\f_i\}_{i = 1}^d$ is a linearly independent basis $\{\f_i\}_{i = 1}^d$ of $\R[\F_X]$ (here  $\dim\left( \R[\F_X] \right) = d$), i.e. it is a set of OPFs  $\{\f_i\}_{i = 1}^d$ such that every other $\f \in \F_X$ can be uniquely expressed as $\f(x) = \sum_{i = 1}^d c_i \f_i(x)$ for all $x \in X$ and $c_i$ coefficients in $\R$. In this representation a state $\Omega_x$ is written as:
\begin{align}
\Omega_x = 
\begin{pmatrix}
\f_1(x) \\
\vdots \\
\f_d(x)
\end{pmatrix} \ ,
\end{align}
and an effect $\Lambda_\f$ (where $\f(x) = \sum_{i = 1}^d c_i \f_i(x))$ is a dual vector:
\begin{align*}
\Lambda_\f = \left( c_1, ..., c_d \right) \ .
\end{align*}
One can immediately verify that $\Lambda_\f \cdot \Omega_x = \f(x)$.
\end{remark}

In general the convex hull of a set of points $P$ will not have that set of points as extremal points, since generically some points in $P$ might lie in the convex hull of other points of $P$. As such it is not immediate that $\conv(\Omega_X)$ has extremal points $\Omega_X$. The following lemma tells us that this is the case. Let us denote by $\delta_e(C)$ the extremal points of some convex set $C$.

\begin{lemma}\label{lem:extremal_points}
The embedding of a system $\S_X = \{X, G , \F_X\}$ into $(V,V^*)$ with maps $\Omega, \Gamma,\Lambda$ is such that $\delta_e\left(\conv\left(\Omega_X\right)\right) = \Omega_X$. 
\end{lemma}

This lemma is proven in Appendix~\ref{app:extremal_points_lem}, and makes use of the transitivity of the action of $G$ on the pure states $X$. It follows from the fact that $\Omega_X$ is a subset of a hypersphere in the affine span of $\Omega_X$ centred on the maximally mixed state. See also~\cite[Proposition 2.2]{Sanyal_2011} for an equivalent statement of the lemma and proof.

\begin{definition}[Tomographically equivalent probabilistic structures]
Two probabilistic structures $\F$ and $\F'$ are tomographically equivalent if they yield the same equivalence classes of ensembles (i.e. mixed states).
\end{definition}

We note that two systems $\S_X = (X,G, \F_X)$ and $\S_X' = (X,G, \F_X')$ with embeddings $\Omega,\Gamma,\Lambda$ and $\Omega', \Gamma',\Lambda'$  are tomographically equivalent if and only if $\Omega_X$ and $\Omega_X'$ are affinely isomorphic (i.e. equivalent as convex set).

For a given system the asymptotic limit consists of the scenario where all preparation procedures are of $n$ copies of the same state and $n$ tends to infinity. In this case all states (including mixed) become perfectly distinguishable (though this does not lift the degeneracy of equivalent ensembles). We denote $\bar \F_X$ the equivalence class of all tomographically equivalent probabilistic structures, hence $\left( X, \bar \F_X \right)$ can be identified with the state space (convex set) $\conv(\Omega_X)$ which is the same for all systems $\left(X, \F_X \right)$ with  $\F_X \in \bar \F_X$. A representative element is the probabilistic structure corresponding to the (effect) unrestricted system.

\begin{remark}[On the link between tomographically equivalent probabilistic structures and restriction of effects]
	The notion of tomographically equivalent probabilistic structures can be cast in terms of \emph{restriction of effects}. A state space is \emph{effect unrestricted} when all linear functionals $\GL(\R[\F_{G/H}]^*) \to [0,1]$ correspond to allowed measurement outcomes. A system is restricted when some of the mathematically allowed functionals do not represent any measurement outcomes of the theory. However when a system has restricted effects, it is always the case that the allowed effects span the dual space $V^*$ of the state space embedded in $V$. In other words both the restricted and unrestricted systems have the same mixed states (the restricted effects are always such that they separate the initial state space). A system with restricted effects has a tomographically equivalent probabilistic structure to the unrestricted system. Two tomographically equivalent probabilistic structures can be obtained by restriction of a common probabilistic structure.
\end{remark}

\subsubsection{Topological and differentiable features of the probabilistic representation}

In the case where the dynamical structure $(X,G)$ has topological/differentiable features, we assume that the maps $\Omega: X \to V$ and $\Gamma: G \to \GL(V)$ are continuous/smooth, implying that the action $\Omega_x \to \Omega_{gx}$ is also continuous/smooth. In the cases where $\F_X$ separates $X$ and $G$ (implying that $\Omega$ and $\Gamma$ are injections) the inverse maps $\Omega^{-1}$ and $\Gamma^{-1}$ are also assumed to be continuous, implying that $\Omega_X$ and $\Gamma_G$ are homeomorphic/diffeomorphic to $X$ and $G$ respectively.  

For topological groups a group representation is a continuous homomorphism $\Gamma: G \to \GL(V)$ (smooth for Lie groups), hence the continuity/smoothness of $\Gamma$ will allow us to make use of the representation theory of topological/Lie groups in the rest of the work.  Since $\R[\F_X]$ is assumed to be finite dimensional continuity of these maps entails that the functions $\f$ are continuous on $X$.

\begin{remark}
The assumption that $\Omega: X \to V$ and $\Gamma: G \to \GL(V)$ are continuous/smooth is justified by the following.  First note that continuity of these maps implies that $\Omega_X$ and $\Gamma_G$ are isomorphic to $X$ and $G$ not just as set/groups, but as topological spaces/topological groups (differentiable manifolds/Lie groups). 
Consider the case where the dynamical structure has topological/differentiable structure but the maps $\Omega$ and $\Gamma$ are not continuous/smooth, i.e. $\Omega_X$ and $\Gamma_G$ are not homeomorphic/diffeomorphic to $X$ and $G$ respectively. Then given access only to the operational system, with state space $\conv(\Omega_X)$ and transformation space $\conv(\Gamma_G)$ and asked to reconstruct the dynamical structure, we would not assign it a dynamical structure with $X$ a topological space acted on continuously by the group $G$ (or differentiable manifold acted on smoothly by a Lie group). Rather we would assign it the set of pure states $X$ without any topological structure. 
To summarise: the operational perspective begins from some experimental data, then constructs the convex state, transformation and effect spaces and only then can one infer the pure states and reversible transformations of those systems (i.e. dynamical structure). From this perspective the dynamical structure has all the structural properties of the experimentally determined $\Omega_X$ and $\Gamma_G$, implying that the maps $\Omega$ and $\Gamma$ must preserve these structures.
\end{remark}

\subsection{Non-compactness and ``Possibility of state estimation using a finite outcome set''}\label{sec:non-comp_dyn_struc}

Although we have restricted our attention to probabilistic systems with $X$ and $G$ compact, there are well defined probabilistic systems which are non-compact such as infinite dimensional quantum systems. These systems violate ``Possibility of state estimation using a finite outcome set'' and a natural question to ask is whether ``Possibility of state estimation using a finite outcome set'' rules out non-compact dynamical structures in general.
In the following we no longer assume $X$ and $G$ compact, and briefly explore this question.
\begin{lemma}~\label{lem:comp_metric}
Under the assumption of ``Possibility of state estimation using a finite outcome set'' the sets $\conv \left(\Omega_X\right)$ and $\conv\left(\Gamma_G\right)$ are compact.
\end{lemma}

Observe however that compactness of $\conv\left(\Omega_X\right)$ and $\conv\left(\Gamma_G\right)$ does not imply compactness of $\Omega_X$ and $\Gamma_G$. There exists compact subset of $\R^n$ whose extremal points are not closed (for an example of such a set see~~\cite[proof of Lemma 0.22]{extreme_libor_2010}). As such one cannot use compactness of a convex set to infer compactness of its extremal points. Although there exist compact convex sets in $\R^n$ with a non-compact set of extremal points, it is not known to the authors whether any such sets where the extremal points are transitive under some group action exist. As such it may be the case that for transitive dynamical structures ``Possibility of state estimation using a finite outcome set'' imposes that $X$ and $G$ are compact.

In the case of transitive non-compact dynamical structures with a non-compact group one can make use of group representation theory to rule out the existence of probabilistic structure compatible with the assumption of ``Possibility of state estimation using a finite outcome set''. For many non-compact groups $G$  (such as non-compact simple Lie groups) it is known that there are no non-trivial finite dimensional continuous unitary representations, which rules out probabilistic structures which violate ``Possibility of state estimation using a finite outcome set'' for dynamical structures with those groups. An open question is whether are there any transitive dynamical structures, where $X$ and $G$ are non-compact, which are consistent with ``Possibility of state estimation using a finite outcome set''.

\section{Classification theorem}\label{sec:classification}

Before stating the main theorem of this section we will need to define the notion of a \emph{Gelfand pair}. Here $(\Gamma, V, \bF)$ refers to a representation $\Gamma: G \mapsto \GL(V)$ over a field $\bF$.  

\begin{definition}[Gelfand pair]
A pair $(G,H)$ with $G$ a group and $H$ a subgroup of $G$ form a Gelfand pair when for all irreducible representations $(\Gamma, V, \C)$ of $G$, the restriction $\Gamma_{|H}$ has at most one trivial sub-representation.
\end{definition}

In other words, for a Gelfand pair $(G,H)$, every irreducible representation $(\Gamma, V, \C)$ of $G$ is such that all the vectors $v \in V$ which are invariant under $H$ span a subspace of dimension at most 1. A vector $v \in V$ is invariant under $H$ when $\Gamma_h v = v$ for all $h \in H$.

This definition applies to complex irreducible representations. For irreducible representations over the field $\R$ the restriction $\Gamma_{|H}$ may contain two trivial sub-representations, however all $H$-invariant vectors are related by invertible transformations which  commute with the group action (this does not contradict Schur's Lemma, which applies to irreducible representation over the complex field). More details and proofs can be found in Appendix~\ref{app:real_rep}. 

A representation $(\Gamma, V, \C)$ of a group $G$ which has a non-zero $H$-invariant vector (i.e. for which $\Gamma_{|H}$ contains a trivial sub-representation) is called a \emph{spherical representation} of $(G,H)$.

\begin{theorem}[Classification theorem]\label{thm:classification}
	Let  $\D = (G,H)$ be a transitive dynamical structure, and let us consider probabilistic structures $\F_{G/H}$ such that $\R[\F_{G/H}]$ is finite-dimensional. By Theorem~\ref{thm:sys_maps} every system $\S_{G/H} = (G,H, \F_{G/H})$ has an associated representation $\Gamma: G \to \GL(\R[\F_{G/H}]^*)$. 
	\begin{enumerate}[label = \roman*.]
		\item Every probabilistic structure $\F_{G/H}$  (up to tomographic equivalence) has an associated representation $\Gamma$ of the form:
		\begin{align}\label{eq:classification}
		\Gamma = \bigoplus_{j} \Gamma_j ,
		\end{align}
		where each term $(\Gamma_j,V_j, \R)$ is a real-irreducible representation with at least one trivial subrepresentation when restricted to $H$.
		\label{thmpart:one}
		\item Conversely every representation of the form \eqref{eq:classification} (where each irreducible representation in the decomposition has at leat one trivial subrepresentation when restricted to $H$) is associated to at least one probabilistic structure $\F_{G/H}$.    \label{thmpart:two}
		\item When $(G,H)$ forms a Gelfand pair the correspondence between representations $(\Gamma, V , \R)$ of the form \eqref{eq:classification} and probabilistic structures (up to tomographic equivalence) $\F_{G/H}$  is one-to-one. \label{thmpart:three}
		\item When $(G,H)$ does not form a Gelfand pair then some representations $(\Gamma, V , \R)$ of the form \eqref{eq:classification} have infinitely-many tomographically inequivalent probabilistic structures $\F_{G/H}$ associated to them. \label{thmpart:four}
	\end{enumerate}
\end{theorem}
This theorem is proven in Appendix~\ref{app:classification_thm}.

Parts \ref{thmpart:one} and \ref{thmpart:three}  entail that for a dynamical structure $(G,H)$ which form a Gelfand pair one can classify all possible probabilistic structures $\F$ (up to equivalence) by finding the irreducible representations $\Gamma$ of $G$ such that $\Gamma_{G|H}$ has a trivial representation.

Parts \ref{thmpart:three} and \ref{thmpart:four} tell us that for  Gelfand pairs all inequivalent probabilistic structures are characterised by different representations of $G$.  Therefore for Gelfand pairs all probabilistic structures are in one-to-one correspondence with representations of the dynamical group, up to restriction of effects. For non-Gelfand pairs there are inequivalent probabilistic structures which are associated to the same representation of $G$.

The one-to-one correspondence between probabilistic structures and representations for Gelfand pairs is a direct consequence of the existence of an invertible transformation which commutes with group action for all invariant $H$-vectors (see Corollary~\ref{cor:commute-H_vec} in Appendix~\ref{app:real_rep}). For real irreducible representations which are also complex irreducible this is just the identity (by Schur's lemma), however for real irreducible of complex type (i.e. which are complex reducible) the linear space of transformations which commutes with all $H$-invariant vectors is two dimensional. As shown in Lemma~\ref{lem:rea_comp_struc} there are no real irreducible representations of quaternionic type which have an $H$-invariant vector when $(G,H)$ Gelfand.

We observe that this theorem does not guarantee that for a given representation $\Gamma$ of the form~\eqref{eq:classification} the associated OPF set $\F$ separates the pure states. For instance the trivial representation $\Gamma: G \to \GL(\R)$, $\Gamma(g)= \I_{\R}$ for all $g \in G$ is such that any vector $v \in \R$ is $H$-invariant, and the state space obtained for any choice of non-zero reference vector $v$ is trivial: $\Omega_x = v$ for all $x \in X$.

\section{Rigidity of dynamical structures}\label{sec:rigid}

In this section we analyse which probabilistic structures can be continuously deformed~\footnote{Here by continuity we do not mean in the sense of a continuous map between topological spaces, but rather in the sense that there exists a connected path between the two probabilistic structures being deformed.}. We first study which dynamical structures $(G,H)$ have probabilistic structures which are arbitrarily close. Following this we show that these probabilistic structures can be continuously deformed.
In order to do so, we define an operational distance between probabilistic structures in terms of how difficult is to discriminate them. 

Obviously, one can always smoothly deform a probabilistic structure by restricting the set of OPFs; for example, by adding noise to the measurements~\footnote{Adding noise to a measurement $\{\Lambda_{\f_i}\}_i$ consists of replacing each effect $\Lambda_{\f_i}$ by $(1-\lambda)\Lambda_{\f_i} +  \lambda \Lambda_\u$ for some  noise parameter $\lambda \in [0,1]$}. However, all these variants have the same set of mixed states, or in other words, the same equivalence classes of ensembles of pure states $\{ (p_i, x_i)\}_i$. We call all these probabilistic structures \emph{tomographically equivalent} because, in estimation processes with multiple measurements, they agree on the set of mixed states.
In each tomographically-equivalent class of probabilistic structures there is a privileged element: the unrestricted probabilistic structure. This $\F$ includes all linear maps $\Lambda: V\to \mathbb R$ that map pure states to probabilities $\Lambda: \Omega(X) \to [0,1]$.
In order to avoid considering trivial deformations (i.e. those which leave the space of mixed states unchanged), in this section, we only consider unrestricted probabilistic structures.

A probabilistic structure $\F_0$ for which every other probabilistic structure $\F_1$ of the same linear dimension is at a finite bounded distance is called \emph{rigid}. In other words, once the dimension of the space of mixed states is fixed, there is a finite bound on the minimal error when discriminating between probabilistic structures compatible with that dimension.

Theorem~\ref{thm:classification} tells us that if a dynamical structure $(G,H)$ is a Gelfand pair then the set of unrestricted probabilistic structures is countable.
We prove that each finite-dimensional probabilistic structure of a Gelfand pair $(G,H)$ is rigid. We show that for non-Gelfand pairs there exists probabilistic structures $\F_0$ which are not rigid, and which can be continuously deformed to other probabilistic structures of the same linear dimension.

\subsection{Distance between inequivalent probabilistic structures}

For a given dynamical structure $\left( G, H \right)$ (with $X \cong G/H$) there is a natural notion of distance between probabilistic structures $\F_X$. The distance between two OPFs $\f^0 \in \F^0_X$ and $\f^1 \in \F^1_X$ is given by:
\begin{align}
\dist(\f^0,\f^1) = \max_{x \in X} |\f^0(x) -\f^1(x)| \ .
\end{align}
This distance is directly related to the minimal error made when discriminating between $\f^0$ and $\f^1$.
We define the distance between two probabilistic structures $\F^0_X$ and $\F^1_X$ as:
\begin{align}
  D(\F^0,\F^1) = \max_{\f^0 \in \F^0_X}  \min_{\f^1 \in \F^1_X} \dist(\f^0,\f^1) \ ,
\end{align}
which informs us about the error that we make when certifying that a system behaves according to $\F^0_X$ and not $\F^1_X$ in the optimal experimental setting $\f^0 \in \F^0_X$. 
Note that $D$ is not symmetric and hence it is not a metric distance. We introduce the symmetrised distance:
\begin{align}\label{eq:dist_prob}
D_{\rm sym}(\F^0,\F^1) = \max \{D(\F^0,\F^1),D(\F^1,\F^0)  \} \ ,
\end{align}
which is a metric distance. The following theorem (proven in Appendix~\ref{app:def_proof}) provides us with a lower bound on the distance between certain pairs of probabilistic structures $\F^0$ and $\F^1$.

\begin{theorem}\label{thm:deformation}
	Let the dynamical structure $(G,H)$ be a Gelfand pair, and let $\F^0$ and $\F^1$ be two unrestricted probabilistic structures of $(G,H)$.
	If $\F^0$ has an irreducible representation of dimension $d^0$ which does not appear in $\F^1$ then 
	\begin{equation}
	D(\F^0,\F^1) \geq \frac 1 {4 d^0} \ .
	\end{equation}
\end{theorem}

Now we recall that for Gelfand systems, two unrestricted probabilistic structures are equal if and only if they have the same irreps in their decomposition.
Hence, the above theorem implies that, for Gelfand systems, each pair of unrestricted probabilistic structure can be discriminated by finite means.

\subsection{Rigid and non-rigid probabilistic structures}

Theorem~\ref{thm:deformation} tells us that, if we fix a dynamical structure consisting of a Gelfand pair $(G,H)$, then the hypothesis that ``the observed data is generated by a particular probabilistic structure $\F_0$ of (G,H)", in opposition to ``the observed data is not generated by $\F_0$", can be tested with finite means.
We now look at the property of rigidity of probabilistic structures, i.e. which probabilistic structures are such that every other probabilistic structure of the same linear dimension is at a finitely bounded distance.

\begin{theorem}\label{thm:deformation2}
Let $\F^0$ of dimension $\dim \R[\F_0] = d_0$ be an unrestricted probabilistic structure of $(G,H)$ with associated representation $\Gamma_G$.
\begin{enumerate}[label = \roman*.]
\item  If every pair of $H$-invariant vectors in $\R[\F^0]$  is related by an invertible transformation which commutes with $\Gamma_G$ then $\F^0$ is rigid and any other inequivalent probabilistic structure $\F_1$ such that $\dim \R[\F_1] = d_0$ is at distance:
	\begin{equation}
	D_{\rm sym} (\F^0,\F^1) \geq \frac 1 {4 (d_0-1)} \ ,
	\end{equation}
\item If there is a pair of  $H$-invariant vectors in $\R[\F^0]$ that are not related by any invertible transformation which commutes with $\Gamma_G$ then $\F^0$ is non-rigid and for any $\epsilon>0$ ($\epsilon \ll 1$) there is an inequivalent probabilistic structure $\F^1$ with $\dim \R[\F_1] = d_0$  at distance
\begin{equation}
D_{\rm sym} (\F^0,\F^1) \leq 2 \epsilon \ .
\end{equation} 
\end{enumerate}
\end{theorem}

This theorem is proven in Appendix~\ref{app:def_prob_struc}. 

For Gelfand pairs all irreducible spherical representations $\Gamma_G$ are such that all pairs of $H$-invariant vectors are related by invertible transformations which commute with $\Gamma_G$, hence all probabilistic structures for Gelfand pairs are rigid. 
For non-Gelfand pairs there exist probabilistic structures $\F^0$ which have associated representations such that for all pairs of $H$-invariant vectors there is no invertible transformation relating them which commutes with $\Gamma_G$. 
Hence we have the following corollary:
\begin{corollary}\label{cor:gelf_rigid}
	Let $\D = \left(G,H \right)$ be a dynamical structure. 
	\begin{enumerate}
		\item If $(G,H)$ is a Gelfand pair, then every unrestricted probabilistic structure $\F_{G/H}$ is rigid. 
		\item If $(G,H)$ is not a Gelfand pair, then there exist probabilistic structures $\F_{G/H}$ which are not rigid, which are those with associated representations $\Gamma_G$ which admit $H$-invariant vectors related by invertible transformations which do not commute with $\Gamma_G$.
	\end{enumerate}
\end{corollary}

In Lemma~\ref{lem:deformation3} we show that these probabilistic structures can be \emph{continuously deformed} to other probabilistic structures of the same linear dimension. Deformation of probabilistic structure is defined in Section~\ref{sec:def_map} where deformation maps between different probabilistic structures are explicitly characterised. 

Before studying the general case we provide an example of a non-rigid probabilistic structure and how to continuously deform it.

\subsection{Continuous deformation of probabilistic structures: an example}\label{sec:def_ex}

In this section we analyse a dynamical system $(G,H)$ that is not a Gelfand pair. Hence, some of its probabilistic structures can be continuously deformed, giving rise to varying statistical properties.
This is an interesting feature of GPTs that has not been explored in the literature.

\begin{definition}[The deformable state space]\label{def:family}
Consider the set of pure states $\Omega_X= \{U\Omega_{x_0}U^\dagger : \forall\, U\in {\rm SU}(3)\}$ generated by the adjoint action of $G = {\rm SU}(3)$ on the reference state
\begin{equation}\label{eq:refstate1}
\Omega_{x_0} 
= 
\left(\begin{array}{ccc}
\alpha_1 & 0 & 0 \\
0 & \alpha_2 & 0 \\
0 & 0 & \alpha_3
\end{array} \right) ,
\end{equation}
where the three real coefficients $\alpha_i$ are different ($\alpha_{i} \neq  \alpha_{j}$ for all $i, j$) and add up to one $\sum_i \alpha_i =1$.   
This state space has stabiliser subgroup 
\begin{equation}\label{eq:stabgr}
  H = 
\left\{ \left(\begin{array}{ccc}
e^{i\phi_1} & 0 & 0 \\
0 & e^{i\phi_2} & 0 \\
0 & 0 & e^{-i(\phi_1+\phi_2)}
\end{array} \right):\, \forall\, \phi_1, \phi_2 \in \R \right\}
  \cong {\rm S}\left(\U(1) \times \U(1) \times \U(1)\right)\ .
\end{equation}
\end{definition}

We observe that the embedding of the pure states is given $\Omega: X \to V$ where $V$ is the real space of $3 \times 3$ Hermitian  matrices. In this representation outcome probabilities are given by effects (i.e.  linear functionals of the states $\Omega_x$), hence they are given by the trace inner product: $\Lambda_\f (\Omega_x) = \tr \left(H_\f \Omega_x\right)$ where $H_\f$ a Hermitian matrix. The unit effect $\Lambda_\u$ is linear functional evaluating to $1$ on all $\Omega_x$, hence $\Lambda_\u(\Omega_x) = \tr(\Omega_x)$.

The case with two equal coefficients is equivalent to the familiar three-level quantum system, $(\alpha_1, \alpha_2, \alpha_3) = (1,0,0)$, which has stabiliser subgroup $H \cong {\rm S}\left(\U(2) \times \U(1) \right)$~\cite{Galley_classification_2017} different than our example \eqref{eq:stabgr}. 
This illustrates how we cannot deform quantum theory without changing its stabiliser subgroup and hence changing its dynamical structure $(G,H)$.
By contrast, the above-defined family of state spaces can be deformed without changing the stabiliser nor the dynamical structure. 

The pair $(\SU(3) ,{\rm S}\left(\U(2) \times \U(1) \right) )$ is a Gelfand pair~\cite{Galley_classification_2017}, whereas $(\SU(3), {\rm S}\left(\U(1) \times \U(1) \times \U(1)\right)$ is not. To show this we need to find a single irreducible representation $\Gamma$ of $G$ such that the restriction $\Gamma_{|H}$ to the subgroup $H$ has more than one trivial subrepresentation. Take the adjoint action of $G$ (see Definition~\ref{def:family}) acting on the full complex space of complex matrices; this representation decomposes into a trivial representation of $G$ (acting on the subspace spanned by the identity matrix) and the adjoint representation acting on the complementary subspace. This subspace is spanned by the trace $0$ complex matrices and carries an irreducible representation of $G \cong \SU(3)$ (the adjoint representation). We observe that all diagonal matrices are invariant under the adjoint action of the subgroup $H$ defined in Equation~\eqref{eq:stabgr}, and as such the space of $H$-invariant vectors is spanned by the trace $0$ diagonal matrices. This is a 2 dimensional subspace of the full space of trace $0$ matrices carrying the irreducible representation of $G$ implying that $(\SU(3), {\rm S}\left(\U(1) \times \U(1) \times \U(1)\right)$ is not a Gelfand pair.
 
We know that the three-level quantum system has three perfectly distinguishable states.
The following theorem tells us that this is not the case for above-defined state spaces.

\begin{theorem} 
All state spaces introduced in Definition~\ref{def:family} have two perfectly distinguishable states and no more.

\end{theorem}

\begin{proof}
In the following proof we write $x$ instead of $\Omega_x$ (similarly $X$ instead of $\Omega_X$).

Let us start by assuming the existence of three perfectly distinguishable states $x_1, x_2, x_3\in V$. This implies the existence of a three-outcome measurement $A_1, A_2, A_3\in V$ such that ${\rm tr}(A_i x_j) = \delta_{ij}$. 
Without loss of generality we can take the three states to be pure $x_i\in X \subseteq V$.
In the following analysis we use a $V$-basis where $A_1$ is diagonal
\begin{equation}\label{eq:refstate2}
A_1 
= 
\left(\begin{array}{ccc}
\gamma_1 & 0 & 0 \\
0 & \gamma_2 & 0 \\
0 & 0 & \gamma_3
\end{array} \right) \ .
\end{equation}
The probability of $A_1$ with any state $x$ only depends on the diagonal of the state (in this basis). Therefore, in what follows, we characterise the projection of ${\rm conv} \left(X \right) \subseteq V$ into the diagonal.
A general state $Ux_0U^\dagger$ has diagonal projection
\begin{equation}
[U x_0 U^\dagger]_{ii}
=
\sum_j U_{ij} \alpha_j [U^\dagger]_{ji}
=
\sum_j |U_{ij}|^2 \alpha_j\ .
\end{equation}
The unitarity of $U$ implies that $|U_{ij}|^2$ is a doubly stochastic matrix, and Birkhoff's theorem tells us that $|U_{ij}|^2$ is a mixture of the six permutation-matrices of three elements.
Conversely, the six permutation matrices can be written as $|U_{ij}|^2$. This, together with the convexity of the state space, implies that the projection of ${\rm conv}X$ into the diagonal is the convex set generated by the six extreme points
\begin{equation}\label{eq:refstate3}
y_\sigma
=
\left(\begin{array}{c}
\alpha_{\sigma(1)} \\
\alpha_{\sigma(2)} \\
\alpha_{\sigma(3)}
\end{array} \right),\ 
\sigma \in \mathcal S_3\ ,
\end{equation}
where $\mathcal S_3$ is the group of permutations of 3 elements.
These six points (also denoted $y_1, y_2,\ldots, y_6$) are depicted in Figure~\ref{fig:hexagon}.

Let us show that each of the three pairs of opposite lines in Figure~\ref{fig:hexagon} are indeed parallel. For example:
\begin{align}
  y_1 -y_2 = \left(\begin{array}{c}
    \alpha_1 \\
    \alpha_2 \\
    \alpha_3
  \end{array} \right)
  - \left(\begin{array}{c}
    \alpha_1 \\
    \alpha_3 \\
    \alpha_2
  \end{array} \right)
  \propto \left(\begin{array}{c}
    0 \\
    1 \\
    -1
  \end{array} \right),
  \\
  y_6 -y_3 = \left(\begin{array}{c}
    \alpha_2 \\
    \alpha_3 \\
    \alpha_1
  \end{array} \right)
  - \left(\begin{array}{c}
    \alpha_2 \\
    \alpha_1 \\
    \alpha_3
  \end{array} \right)
  \propto \left(\begin{array}{c}
    0 \\
    1 \\
    -1
  \end{array} \right),
  \\
  y_5 -y_4 = \left(\begin{array}{c}
    \alpha_3 \\
    \alpha_1 \\
    \alpha_2
  \end{array} \right)
  - \left(\begin{array}{c}
    \alpha_3 \\
    \alpha_2 \\
    \alpha_1
  \end{array} \right)
  \propto \left(\begin{array}{c}
    0 \\
    1 \\
    -1
  \end{array} \right).
\end{align}

\begin{figure}[t]
	\includegraphics{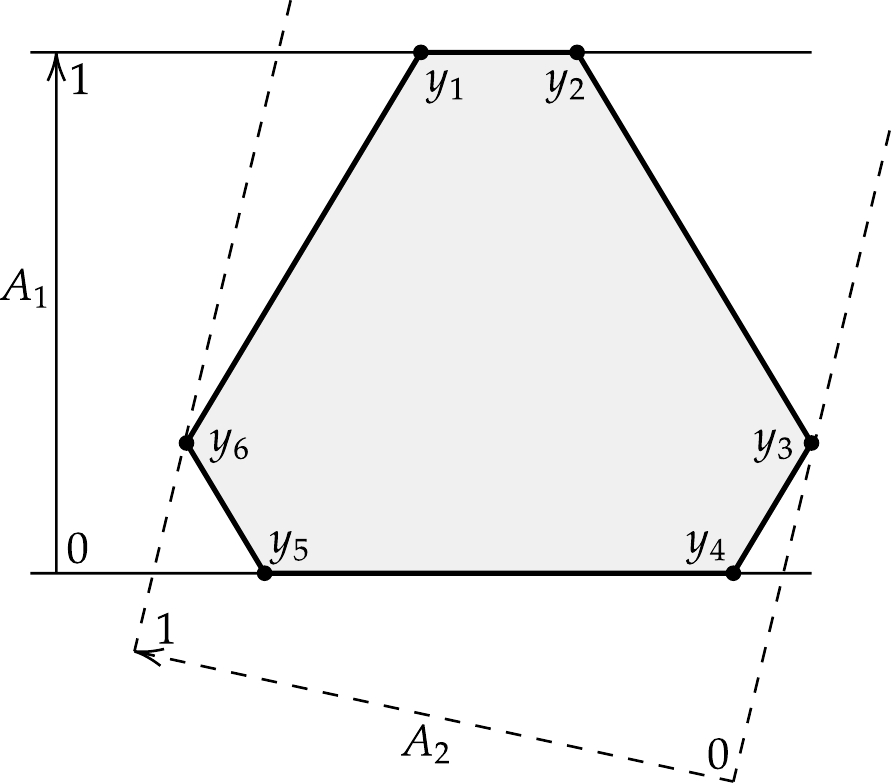}
	\centering
	\caption{This six-sided figure is the projection onto the diagonal of a generic state space of the family introduced in Definition~\ref{def:family}. 
	The extreme points $y_1,\ldots , y_6$ are the six permutations of the vector $(\alpha_1, \alpha_2, \alpha_3)$, which can be represented in a two-dimensional plane because of the normalisation condition $\sum_i \alpha_i =1$.
	In the generic case $(\alpha_1, \alpha_2, \alpha_3)$ the figure has two types of sides with alternating length, and for certain values of $(\alpha_1, \alpha_2, \alpha_3)$ the six sides have equal length. 
	In the quantum case $(\alpha_1, \alpha_2, \alpha_3) = (1,0,0)$ the short sides have zero length zero and the figure becomes a triangle.
	The outcome $A_1$ ($A_2$) has probability 0 or 1 in the solid (dotted) lines.
}\label{fig:hexagon}
\end{figure}

Condition ${\rm tr}(A_1 x_j) = \delta_{1j}$ implies that the scalar product $(\gamma_1, \gamma_2, \gamma_3) \cdot y_\sigma$ takes the value zero for two permutations $\sigma$ and the value 1 on at least one permutation. 
However, as shown in Figure~\ref{fig:hexagon}, the only outcome that tells apart states $y_1$ from $y_4, y_5$ is $A_1$, which gives probability one for states $y_1, y_2$ and zero for $y_4, y_5$. 
It is worth mentioning that the vectors $y_1, y_2, y_4, y_5$ correspond to four pure states with zero off-diagonal components. Hence, the outcome probabilities of these pure states can be calculated by only looking at the diagonal projection (i.e.~the figure).

The figure also shows that, no matter how we choose the direction $A_2$, the states $y_4, y_5$ (or $y_1, y_2$) cannot be perfectly distinguished. 
This proves the non-existence of three perfectly distinguishable states in this family of state spaces.
Of course, this argument breaks down for the three-level quantum system, when the projection becomes a triangle instead of an hexagon.
\end{proof}

The state spaces under consideration (Definition~\ref{def:family}) have a remarkable property that is not present in quantum theory.
This property is sometimes called ``violation of no-simultaneous encoding"~\cite{Masanes_existence_2013} and it is very similar to ``information causality" \cite{IC}.
This property allows to perfectly encode one bit of information (e.g.~$y_1, y_2$ versus $y_4, y_5$) and simultaneously imperfectly encode another bit ($y_1, y_5$ versus $y_2, y_4$). Although only one of the two bits can be retrieved, there is a sense in which this system encodes more than one bit of information despite having only two perfectly-distinguishable states.  
Different choices of $(\alpha_1, \alpha_2, \alpha_3)$ will give different success probability when optimally guessing the second bit. This is a statistical feature that distinguishes inequivalent values of $(\alpha_1, \alpha_2, \alpha_3)$ within the family of state spaces of Definition \ref{def:family}.

\subsection{Continuous deformation of probabilistic structures: the general case}\label{sec:def_map}

In the following we only consider systems up to tomographic equivalence, where two systems are tomographically equivalent if and and only if they have the same equivalence classes of ensembles (i.e. the same mixed states). For a given set of pure states $X$ each equivalence class of probabilistic structures $\bar \F_X$ induces a map $\Omega: X \to V$ (where $V \cong \R^n$) by Theorem~\ref{thm:sys_maps}. We will define a deformation of a probabilistic structure for $X$ as a map between different images $\Omega_X$ of $\Omega$-maps for $X$ (since there is a one-to-one correspondence between equivalence classes of probabilistic structures $\bar \F_X$ and images $\Omega_X$ of $\Omega$-maps) satisfying certain conditions which we will formalize with the aid of some further definitions. We only consider equivalence classes of probabilistic structures which separate $X$.

For a given dynamical structure $(X,G)$ let us call $\mathfrak F$  the space of all images $\Omega_X$  (equivalently the space of equivalence classes of  probabilistic structures $\bar \F_X$). For a given $d \in \R^+_{\0}$ let us call $\mathfrak F_d$ the space of all $\Omega_X$ whose linear span is isomorphic to $\R^d$  (equivalently the space of equivalence classes of  probabilistic structures $\bar \F_X$ such that $\R[\bar \F_X] \cong \R^d$). We observe that $\fF_d$ might be the empty set (if there are no representations of $G$ of the form of Equation~\eqref{eq:classification} acting on $\R^d$). For a Gelfand dynamical structure all $\fF_d$ are countable sets, whilst for non-Gelfand pairs there exist $\fF_d$ which are uncountable.

The symmetrised distance defined in Equation~\eqref{eq:dist_prob} turns $\fF_d$ into a metric space. This metric induces a topology on $\fF_d$. 

We are interested in continuously deforming probabilistic structures, namely how to continuously vary one set of mixed states into another (i.e. one probabilistic structure into another) whilst keeping the linear dimension fixed. First we define a general transformation between sets of embedded pure states $\Omega_X$.

\begin{definition}[Trivial $\Omega$-transformation map]
	For two systems $\S^0_X = \left(X,\bar \F_X^0 \right)$ and  $\S^1_X = \left(X,\bar \F_X^1 \right)$ with associated maps $\Omega^0: X \to V^0$ and $\Omega^1: X \to V^1$ we define the trivial $\Omega$-transformation map $M_{0 \to 1}: \fF \to \fF$, $M_{0 \to 1}: \Omega_X^0 \mapsto \Omega_X^1$  (where it acts as $M_{0 \to 1}:  \Omega_x^0 \mapsto \Omega_x^1$, $\forall x \in X$).
\end{definition}

These trivial $\Omega$-transformation maps allow us to map between any two probabilistic structures in $\fF$. A deformation of probabilistic structure is a specific map between probabilistic structures such that the linear dimension of the space of mixed states is unchanged.

\begin{definition}[Deformation map]
	For two systems $\S^0_X = \left(X,\bar \F_X^0 \right)$ and  $\S^1_X = \left(X,\bar \F_X^1 \right)$ with associated maps $\Omega^0: X \to V^0$ and $\Omega^1: X \to V^1$ where $V^0 \cong V^1 \cong \R_d$ a trivial $\Omega$-transformation map $M_{0 \to 1}: \fF_d \to \fF_d$, $M_{0 \to 1}: \Omega_X^0 \mapsto \Omega_X^1$ is a deformation map.
\end{definition}

\begin{lemma}\label{lem:non-linear_def}
A non-trivial deformation map $M_{0 \to 1}: \Omega_X^0 \mapsto \Omega_X^1$ ($\Omega_X^0 \neq \Omega_X^1$) cannot be extended to a linear transformation on $\spann_\R(\Omega_X^0 )$.
\end{lemma}

\begin{proof}
Since $\bar \F_X^0 $ and $\bar \F_X^1$ separate $X$, $\Omega_X^0$ and $\Omega_X^1$ are homeomorphic.  Let us prove by contradiction and assume that $M_{0 \to 1}$ is linear when extended to $\spann_\R(\Omega_X^0)$. Let us call  $\tilde M_{0 \to 1}: \spann_\R(\Omega_X^0) \to \spann_\R(\Omega_X^1)$ this linear extension of $M_{0 \to 1}$. We observe that $\tilde M_{0 \to 1}$ has a trivial kernel since  no element of $\Omega_X^0$ is in the kernel, hence no element of $\spann_\R(\Omega_X^0)$ is either. Moreover the image of $\tilde M_{0 \to 1}$ is $\spann_\R(\Omega_X^1)$:  any element in $\spann_\R(\Omega_X^1)$ is a linear combination $\sum_i \alpha_i \Omega_{x_i}^1$ of elements in $\Omega_X^1$, and hence by linearity of $\tilde M_{0 \to 1}$  $\sum_i \alpha_i \Omega_{x_i}^1$ is the image under $\tilde M_{0 \to 1}$ of the linear combination $\sum_i \alpha_i \Omega_{x_i}^0$ in the domain $\spann_\R(\Omega_X^0)$. Hence $\tilde M_{0 \to 1}$ is an invertible linear transformation which takes $\conv(\Omega_X^0)$ to $\conv(\Omega_X^1)$ and $\conv(\Omega_X^0) \cong \conv(\Omega_X^1)$. This contradicts the assumption that $\Omega^0$ and $\Omega^1$ are induced by tomographically inequivalent probabilistic structures.
\end{proof}

A non-trivial deformation map $M_{0 \to 1}$ cannot be extended to the spaces of mixed states $\conv\left(\Omega^0_X\right)$ and $\conv\left(\Omega^1_X\right)$, which are not isomorphic as convex sets. For such systems one can use the map $M_{0 \to 1}$ to deform the probabilistic structure $\F^0_X$ to $\F^1_X$ to change the space of mixed  states from $\conv\left(\Omega^0_X\right)$ to $\conv\left(\Omega^1_X\right)$ whilst preserving the linear dimension.

We observe that if we consider the general trivial $\Omega$-transformation maps (i.e. the linear dimension of the two state spaces is no longer required to be the same) then some of these maps do linearly extend to $\spann_\R(\Omega^0_X)$ when $\Omega_X^0 \neq \Omega_X^1$. An example of such a map is the projection from a system $\S^0_{X}$ with an associated representation $\Gamma^0$ which is reducible to a system $\S^1_{X}$ with an associated representation $\Gamma^1$ which is a sub-representation of $\Gamma^0$, i.e. $\Gamma^0 \cong \Gamma^1 \oplus \Gamma^2$ for some representation $\Gamma^3$ of $G$.

\begin{definition}[Continuous deformation of probabilistic structure]
A probabilistic structure $\bar \F_X^0$ can be continuously deformed to another probabilistic structure $\bar \F_X^1$ if there is a continuous function $\gamma: [0,1] \to \fF_d$, $\gamma(t) = \Omega^t_X$ in $\fF_d$ with $t \in [0,1]$ such that $\gamma(0) =  \Omega^0_X$ and $\gamma(1) =\Omega^1_X$. Here continuity is defined with respect to the usual topology on $[0,1]\subset \R$. $\gamma$ defines a connected path in $\fF_d$.
\end{definition}

We observe that the connected path $\gamma(t)$ defines a family of deformation maps $M_{0 \to t}: \Omega^0_X \mapsto \Omega^t_X$ for every $t \in [0,1]$. 

In other words $\Omega_X^0$ can be continuously deformed to $\Omega^1_X$ if there exists a connected path between the two passing through a continuum of different $\Omega_X^t$ (each spanning a linear space of the same dimension). If there is no such path between the two then $\Omega_X^0$ cannot be continuously deformed to $\Omega^1_X$.

In the following theorem we show the relation between the rigidity of a probabilistic structure and the possibility of continuously deforming a probabilistic structure.

\begin{lemma}\label{lem:deformation3}
	Any non-rigid probabilistic structure $\bar \F^0_{G/H}$ can be continuously deformed into another (non-rigid) probabilistic structure $\bar \F_1$. Any rigid probabilistic structure $\bar \F^0_{G/H}$ cannot be continuously deformed to another probabilistic structure.
\end{lemma}

This lemma is proven in Appendix~\ref{app:deformation3}. Together with Corollary~\ref{cor:gelf_rigid} this lemma implies the following corollary.

\begin{corollary}
If $(G,H)$ is a Gelfand pair, then no probabilistic structure can be continuously deformed. If $(G,H)$ is a non-Gelfand pair then it has probabilistic structures which can be continuously deformed.
\end{corollary}

We now provide an explicit construction of a continuous deformation of a probabilistic structure whose associated representation has inequivalent $H$-invariant vectors; for an arbitrary non-Gelfand pair $(G,H)$.

\begin{example}[Continuous deformation]
Consider a non-Gelfand pair $(G,H)$, and an irreducible spherical representation $\Gamma$ acting on $V \cong \R^d$ with inequivalent $H$-invariant vectors $v_H^0$ and $v_H^1$ (which always exists for a given non-Gelfand pair). Any vector in $\spann (v_H^0, v_H^1)$ is $H$-invariant, moreover any two vectors in $\spann (v_H^0, v_H^1)$  which are not proportional will generate inequivalent state spaces under $\Gamma_G$.  Let us take two inequivalent normalised $H$-invariant vectors $v_H^0$ and $v_H^1$. Let us call $\Omega_X^0 = \{\Gamma_g v_H^0| g \in G\}$ and $\Omega_X^1 = \{\Gamma_g v_H^1| g \in G\}$ the two embedded sets of pure states generated from these two reference vectors.  Consider a rotation $R \in \GL (\spann (v_H^0, v_H^1))$ such that $Rv_H^0 = v_H^1$. For a given state $v_{gH}^0 = \Gamma_g v_H^0 \in \Omega_X^0$ and $v_{gH}^1 = \Gamma_g v_{H}^1 \in \Omega_X^1$ we have the relation:
\begin{align}
v_{gH}^1 = \Gamma_g R v_{H}^0 = \Gamma_g R \Gamma_{g^{-1}} v_{gH}^0 \ .
\end{align}
Hence there is a map $m(g) = \Gamma_g R \Gamma_{g^{-1}}$ between the two points $v_{gH}^0 \in \Omega^0_{X}$ and $v_{gH}^1 \in \Omega^1_{X}$, for all points $gH \in X\cong G/H$. The deformation map is then:
\begin{align}
M_{0 \to 1}&: \Omega_X^0 \to \Omega_X^1 \ , \\
M_{0 \to 1}&: v_{gH}^0 \mapsto m(g) v_{gH}^0 =  v_{gH}^1 \ .
\end{align}
Since this map is $g$-dependent (i.e. its action depends on each extremal point $v_{gH}^0$) it is not linear and cannot be extended to the mixed states. 

Let us define $\Omega_X^t$ to be the state space generated by the $H$-invariant reference vector $v_H^t = R(t) v_H^0$ for $t \in [0,1]$ and $m^t(g) = \Gamma_g R(t) \Gamma_{g^{-1}}$. Then the deformation map from $\Omega_X^0$ to $\Omega_X^t$ is given by:
\begin{align}
M_{0 \to t}&: \Omega_X^0 \to \Omega_X^t \ , \\
M_{0 \to t}&: v_{gH}^0 \mapsto m^t(g) v_{gH}^0 =  v_{gH}^t \ .
\end{align}
The family of deformation maps $M_{0 \to t}$ (for $t \in [0,1]$) defines a connected path $\gamma(t) = M_{0 \to t} \Omega^0_X = \Omega_X^t$ in $\fF_d$ (see the proof of Lemma~\ref{lem:deformation3} in Appendix~\ref{app:deformation3}). Hence the probabilistic structure associated to $\Omega^0_X$ can be continuously deformed to the probabilistic structure associated to  $\Omega^1_X$.
\end{example}

\begin{remark}
The possibility of continuously deforming a probabilistic structure without altering its dynamical structure (and without restricting effects) is a very peculiar feature that is not found in any of the known GPTs (such as boxworld and quantum theory over the field of reals, complex or quaternions) to the best of our knowledge.
Moreover we posit that this is a typical feature of GPT systems, in that, most dynamical structures $(G,H)$ are not Gelfand pairs.
If a probabilistic structure can be continuously deformed then the probabilities can be fine-tuned to suitably describe the observed statistics, and hence, make the theory more difficult to falsify.
Hence, the fact that the probabilistic structure of a theory cannot be smoothly deformed makes the falsifiability of the theory more straightforward. 
We believe that this is a desirable property of a theory.
If we consider a dynamical structure $(G,H)$ being a Gelfand pair then we can be sure that any of its probabilistic structures will be straightforwardly falsifiable. 
Finally, it is important to mention that a dynamical structure $(G,H)$ cannot be continuously deformed due to the group and sub-group structures of $G$ and $H$. That is, adding a single element to $G$ or $H$ will generate lots of new elements via products and inverses. And hence, the probabilistic structure is the only part of a theory that, a priori, could be continuously deformed. 
\end{remark}

\section{Gelfand pairs and two point homogeneity}\label{sec:twopoint}

In Theorem~\ref{thm:classification} we have singled out dynamical structures corresponding to Gelfand pairs as being of interest, namely for the convenient property that their probabilistic structures can be classified via the associated group representations. 
This implies that there are countably many of them, and that they cannot be continuously deformed.
This rigidity is a highly desirable property for a fundamental theory of physics, because it does not allow for \emph{ad hoc} parameter adjustment, and is thereby easier to falsify.
Apart from this, one may also ask whether there are other informational/physical motivations for considering Gelfand pairs. One such reason may be the following.
\begin{definition}[Two-point homogeneous action~\cite{Wang_two_1952}]
	A group $G$ acts two-point homogeneously on a metric space $(X,{\rm dist})$ if for every pair of points $(x_1,x_2)$ and $(x_1',x_2')$ in $X$ with ${\rm dist}(x_1,x_2) = {\rm dist}(x_1',x_2')$ there is an element $g \in G$ such that $g x_1 = x_1'$ and $g x_2 = x_2'$.
\end{definition}

Two-point homogeneity implies transitivity, since for any points $x_1$ and $x_2$ we have ${\rm dist}(x_1,x_1) = {\rm dist}(x_2,x_2)$ and hence there exists an element such that $g x_1 = x_2$.
The following is a very remarkable result.

\begin{lemma}[Prop 2.2~\cite{gross_some_1991}]
	If $G$ acts two-point homogeneously on a metric space $X$ and $H$ is the stabilizer of a point, then $(G, H)$ is a Gelfand pair.
\end{lemma}

The requirement of two-point homogeneity restricts us to dynamical structures corresponding to Gelfand pairs. We observe that this requirement requires an additional metric structure to be imposed on the dynamical structure.  
A natural metric on GPT state spaces is the following.

\begin{lemma}{(Pure-state metric distance.)}
The distance between any pair of pure states $x,x' \in X$ defined by
\begin{equation}
  {\rm dist}(x,x')
  =
  \sup_{\f \in \F_X}
  \left[\f(x) -\f(x') \right]
\end{equation}
is bounded ${\rm dist}(x,x') \leq 1$, it satisfies the metric axioms:
\begin{enumerate}
  \item ${\rm dist}(x,x') = 0$ if and only if $x=x'$,
  \item ${\rm dist}(x,x') = {\rm dist}(x',x)$,
  \item ${\rm dist}(x,x'') \leq {\rm dist}(x,x') + {\rm dist}(x',x'')$,
\end{enumerate}
and it is $G$-invariant
\begin{enumerate}
  \item[4.] ${\rm dist}(gx,gx') = {\rm dist}(x,x')$ for all $g\in G$.
\end{enumerate}
\end{lemma}

Therefore we conclude that two-point homogeneous state spaces are Gelfand pairs. It is remarkable that the purely dynamical property of  two-point homogeneity implies that all probabilistic structures are rigid.

However we note that not all Gelfand pairs $(G,H)$ give rise to a homogeneous space $X \cong G/H$ which is  two-point homogeneous. Indeed the classification of all the compact and connected two point homogeneous symmetric spaces was given in ~\cite{Wang_two_1952}. These are listed in Table~\ref{tab:twopoint_hom}.

\begin{table}
	\centering
\begin{tabular}{|c|c|}
	\hline 
$X$	&  $G/H$   \\ 
	\hline 
$S^d$ 	&  $\SO(d)/\SO(d-1)$  \\ 
	\hline 
$\PR^d = S^d/\{\pm I \}$	&  $\O(d)/(\O(d-1) \times \O(1)) \cong \SO(d)/{\rm S}(\O(d-1) \times \O(1))$ \\ 
	\hline 
$\PC^d$	&  $\SU(d)/{\rm S}(\U(d-1) \times \U(1)) \cong \U(d)/(\U(d-1) \times \U(1)$\\ 
	\hline 
$\PH^d$	&  $\Sp(d)/(\Sp(d-1) \times \Sp(1))$\\ 
	\hline  
$\P \bO^3$ & ${\rm F}(4)/\Sp(9)$ \\
	\hline
\end{tabular}
\caption{List of all two-point homogeneous spaces which are connected and compact. Here $\P \bO^3$ is the projective space of octonionic planes known as the Cayley plane. }\label{tab:twopoint_hom}
\end{table}

The full classification of all finite dimensional probabilistic structures for the compact connected two point homogeneous spaces $G/H$ (where all pairs $(G,H)$ corresponding to such spaces are given in Table~\ref{tab:twopoint_hom}) directly follows from the classification of all irreducible spherical representations. Equivalently these are the irreducible subspaces of the function space $C(G/H, \C)$ (continuous functions from $G/H$ to $\C$) under the action of $G$, where a specific basis for an irreducible subspace is given by \emph{spherical harmonics}. This is a generalisation of the well known spherical harmonics for $L^2(S^2)$, where the irreducible representation labelled by $l$ has a basis $Y_{lm}(\theta,\phi)$ spanning a $2l+1$ dimensional subspace.

The $(G,H)$ spherical irreducible representations for these pairs are characterised by a condition on the highest weights given by the \emph{Cartan-Helgason Theorem}~\cite{cartan_sur_1929,helgason_groups_1984}(see~\cite[Theorem 11.4.10.]{wolf_harmonic_2007}. Explicit characterisations of these $(G,H)$ spherical irreducible representations (either in terms of the highest weights or other methods) for the pairs in Table~\ref{tab:twopoint_hom} can be found in the literature.

\section{Grassmannian systems}\label{sec:grassmann}

In this section we introduce a family of non-classical systems which generalise the dynamical structure of quantum systems, and make use of Theorem~\ref{thm:classification} to provide a full classification of these systems. 

The pure states of finite dimensional quantum systems are given by $\mathrm P \mathbb C^d$. This is the set of all one dimensional subspaces of $\mathbb C^d$. We now consider systems with pure states given by the set of all $k$-dimensional subspaces $W \subseteq \mathbb C^d$. This set is known as a Grassmann manifold $\mathrm{Gr}(k,\mathbb{C}^d)$:
\begin{equation}
\mathrm{Gr}(k,\mathbb{C}^d) = \{W \subseteq \mathbb{C}^d , \mathrm{dim}(W) = k\} \ .
\end{equation}
Hence $\PC^d \cong \mathrm{Gr}(1,\mathbb{C}^d )$.
Since $\SU(d)$ acts transitively on $\mathrm{Gr}(k,\mathbb{C}^d )$ it can also be expressed as follows (re-parametrising  $k = m$ and $d = m+n$):
\begin{equation}
\mathrm{Gr}(m,\mathbb{C}^{m+n}) \cong \mathrm{SU}(m+n) / \mathrm{S}(\mathrm{U}(m) \times \mathrm{U}(n))
\end{equation}
Here the embedding of $\mathrm{S}(\mathrm{U}(m) \times \mathrm{U}(n))$ into $\SU(m+n)$ is the direct sum embedding:
\begin{equation}
\mathrm{S}(\mathrm{U}(m) \times \mathrm{U}(n))
=
\left\{ \left(\begin{array}{cc}
e^{i\theta} A & 0 \\
0 & B
\end{array}\right): A\in \mathrm{U}(m),\ 
B\in \mathrm{U}(n),\ 
e^{i\theta m} \det A \det B =1  \right\}\ .
\end{equation}

Similarly one can define Grassmann manifolds over $\R$ and $\H$, generalising the dynamical structures of quantum theory over $\R$ and $\H$. These are:
\begin{align}
\mathrm{Gr}(m,\R^{m+n}) \cong \mathrm{SO}(m+n) / \mathrm{S}(\mathrm{O}(m) \times \mathrm{O}(n)) \ , \\
\mathrm{Gr}(m,\H^{m+n}) \cong \mathrm{Sp}(m+n) / (\mathrm{Sp}(m) \times \mathrm{Sp}(n)) \ .
\end{align}
In the next section we will make use of Theorem~\ref{thm:classification} to classify all possible probabilistic structures for each dynamical structure which is a complex Grassmann manifold.

\subsection{Full classification of all probabilistic structures for complex Grassmann manifolds}

Theorem~\ref{thm:classification} states that for a dynamical structure $(G,H)$ corresponding to a Gelfand pair, every probabilistic structure is in one-to-one correspondence with a spherical representation of $(G,H)$. Hence the first step in classifying probabilistic structures for the Grassmann dynamical structure $\mathrm{Gr}(m,\mathbb{C}^{m+n}) \cong \mathrm{SU}(m+n) / \mathrm{S}(\mathrm{U}(m) \times \mathrm{U}(n))$ is to determine whether $\left(\mathrm{SU}(m+n) , \mathrm{S}(\mathrm{U}(m) \times \mathrm{U}(n)) \right)$ form a Gelfand pair.

\begin{lemma}\label{lem:rep_thry_ex} \begin{minipage}[t]{\linegoal}
	\begin{enumerate}
		\item $(\mathrm{SU}(m+n), \mathrm{S}(\mathrm{U}(m) \times \mathrm{U}(n))$ form a Gelfand pair.
		\item The spherical representations $(\mathrm{SU}(m+n), \mathrm{S}(\mathrm{U}(m) \times \mathrm{U}(n))$ have a real structure. 
	\end{enumerate}
\end{minipage}
\end{lemma}

The first part of the lemma is found in~\cite[Corollary 3]{halima_branching_2007} and the second part is proven in Appendix~\ref{app:rep_thry_ex}. This lemma entails (using Theorem~\ref{thm:classification}) that all probabilistic structures $\F_X$ where $X \cong \mathrm{SU}(n+m)/\mathrm{S}(\mathrm{U}(m) \times \mathrm{U}(n))$ are in one-to-one correspondance with the spherical representations $(\mathrm{SU}(m+n), \mathrm{S}(\mathrm{U}(m) \times \mathrm{U}(n))$. Irreducible spherical representations are typically defined over $\C$, and in general the irreducible representations of a group $G$ over $\C$ are not in one-to-one correspondence with those over $\R$. Part 2. of the lemma allows us to classify the real irreducible spherical representations of $\left(\mathrm{SU}(m+n) , \mathrm{S}(\mathrm{U}(m) \times \mathrm{U}(n)) \right)$ by studying the irreducible spherical representations over $\C$. 

The restriction of representations of $\mathrm{SU}(m+n)$ to  $\mathrm{S}(\mathrm{U}(m) \times \mathrm{U}(n))$ has been studied in \cite{halima_branching_2007}. We summarise the result below. Representations of $\SU(m+n)$ are labelled by a partition $\lambda$ of an integer $k$ in $m+n-1$ parts (often represented as a \emph{Young diagram}). One can construct the associated irreducible representation by applying the \emph{Schur functor} $\mathbb S_\lambda$ to $(\C^{m+n})$~\cite[Proposition 15.15]{Fulton_representation_1991}

\begin{lemma}
The representations of $\mathrm{SU}(n+m)$ which have a trivial representation when restricted to  $\mathrm{S}(\mathrm{U}(n) \times \mathrm{U}(m))$ are of the following form:
When $m =n$:
\begin{equation}\label{eq:rep_one}
\lambda = (2 b _1 , b_1 + b_2 , ... , b_1 + b_m, b_1 - b_m , ... , b_1-b_2,0) \ .
\end{equation}
When $n \geq m + 1 $:
\begin{equation}\label{eq:rep_two}
\lambda = (2 b _1 , b_1 + b_2 , ... , b_1 + b_m, \underbrace{b_1, .... , b_1}_{   n - m \ {\rm times }},  b_1 - b_m , ... , b_1-b_2,0) \ .
\end{equation}
In each case $b_1 \geq b_2 \geq ... \geq b_m \geq 0$. We have added a redundant 0 entry; with it $\lambda$ has length $m+n$. 
\end{lemma}

\subsection{Quartic quantum theory over $\R,\C$ and $\H$}

Quartic quantum theory over $\C$, introduced in~\cite{Zyczkowski_quartic_2008}, is a theory which contains some of the systems classified above. In this theory systems $\S_{k,\C}^{\Quart}$ ($k \in \Z, k >2$) have pure states given by the Grassman manifold $\mathrm{Gr}(k,\mathbb{C}^{k^2})$ and a probabilistic structure $\F^{\tt Quart}_{k,\C}$ given by the adjoint representation. For example the state space for the system $k =2$ can be generated by taking reference state:
\begin{align}
\rho = 
\begin{pmatrix}
1 & 0 & 0 & 0 \\
0 & 1 & 0 & 0 \\
0 & 0 & 0 & 0 \\
0 & 0 & 0 & 0 \\
\end{pmatrix} \ ,
\end{align}
applying the $\SU(4)$ dynamical group in the adjoint representation:
\begin{equation}
U \rho U^\dagger \ ,
\end{equation}
and taking the convex hull of the $\Gr(2,\C^4)$ manifold embedded in $V = {\rm Herm}\left(\C^4\right)$ (the real linear space of Hermitian matrices on $\C^4$).

One problematic feature of quartic quantum theory is that it does not have well defined composition~\cite{Zyczkowski_quartic_2008, Lee_higher_2017}, and as such is just a collection of systems rather than a full theory. 

We can similarly introduce two theories (without composition): \emph{real quartic quantum theory} and \emph{quaternionic quartic quantum theory} where systems are given by $\S_{k,\R}^{\Quart} := \left(\mathrm{Gr}(k,\R^{k^2}), \F^{\tt Quart}_{k,\R}\right)$ and  $\S_{k,\H}^{\Quart} = \left(\mathrm{Gr}(k,\H^{k^2}), \F^{\tt Quart}_{k,\H}\right)$, where  $\R[\F^{\tt Quart}_{k,\R}]$ and $\R[\F^{\tt Quart}_{k,\H}]$ are acted on by the adjoint representation of $\SO(k^2)$ and $\Sp(k^2)$ respectively. In both cases the states space for the system associated to $k = 2$ can be generated by taking the reference state $\rho$ above acting with the adjoint representation of the dynamical group and taking the convex hull.

\section{Discussion}~\label{sec:discussion}

\subsection{Comparison to previous work}

\subsubsection{The generalised quantum mechanics of Mielnik}

The OPF framework presented in this work is similar to the `Group theoretical model' of~\cite{Mielnik_generalized_1974}. The novel aspects of this work include Theorem~\ref{thm:classification} which, building on the framework, establishes a correspondence between probabilistic structures and group representations. We find specific conditions on transitive dynamical structures which make this correspondence one-to-one (namely that the dynamical group and stabilizer subgroup form a Gelfand pair). Moreover Mielnik studies examples with the same pure states as quantum theory, but different dynamical groups. We study (and classify) systems which have different pure states and dynamics.

\subsubsection{Classification of all alternatives to the measurement postulates of quantum system}

Theorem~\ref{thm:classification} is a generalisation of the classification theorem of~\cite{Galley_classification_2017}, where the dynamical structure is no longer constrained to be that of quantum systems. We also find the necessary and sufficient conditions for which dynamical structures have probabilistic structures which are in one-to-one correspondence with group representations.

\subsection{Mapping the space of GPT systems}

In the GPT formalism a \emph{theory} is considered to be a set of systems together with some composition rules. Quantum theory for example is the set of systems ${\rm QT}_\C := \{\S^\Quant_d\}_{d = 2}^{\inf}$ together with the standard tensor product composition rule and partial trace. We note that ${\rm QT}_\C$ alone is not a theory, just a set of systems. 

In this work we also consider sets of systems which are not expected to form theories; these are sets of systems which share a common dynamical structure. For example the set of systems with shared dynamical structure $\left(\PC^2,\SU(2)\right)$ form a \emph{sub-family} of systems, and the set of systems which contain all systems with dynamical structure $\left(\PC^d,\SU(d)\right)$ ($d>2$) form a \emph{family} of systems. 

\begin{figure}
	\centering
	\includegraphics[width=0.7\textwidth]{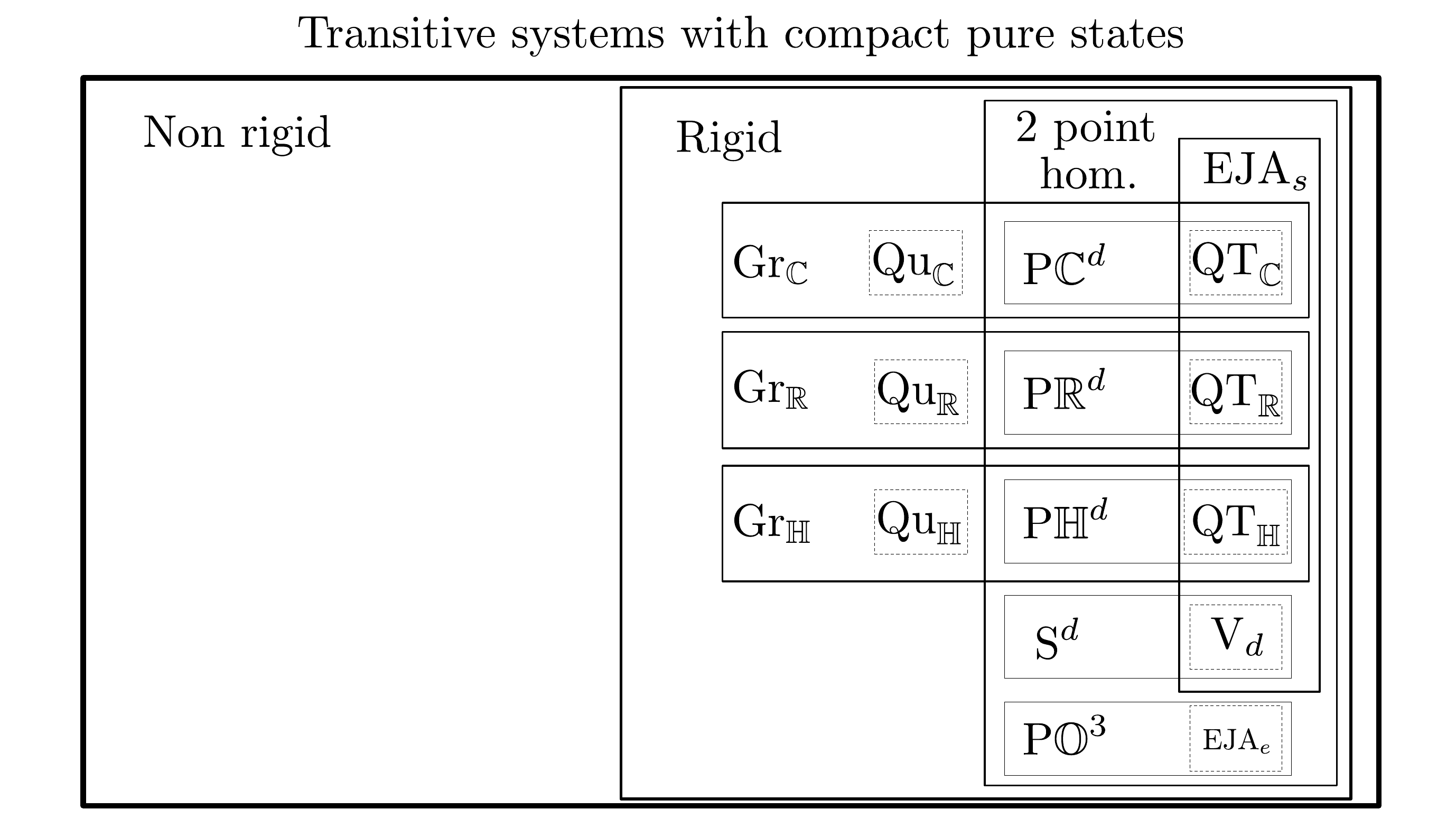}
	\caption{Map of the space of transitive non-classical systems with compact pure states $X = G/H$. `Rigid' and `Non rigid' are notions defined in this paper. `2 point hom.' stands for two point homogeneous. For a field $\bF$,  ${\rm Gr}_\bF$ is the family of systems with pure states given by the Grassmann manifold ${\rm Gr}(\bF^d, \bF^{k})$ for all $2<d<\infty, \ k<d$. ${\rm P}\bF^d$ is the  family of systems with pure states given by projective space over $\bF^d$ for all $1<d<\infty$, hence $\P\bF^d := {\rm Gr}(\bF^d, \bF^{1})$. ${\rm QT}_\bF$ is quantum theory over $\bF$ whilst ${\rm Qu}_\bF$ is quartic quantum theory over $\bF$.  `${\rm EJA}_s$' labels special Euclidean Jordan Algebras (EJA) and `${\rm EJA}_e$' the exceptional EJA. $V_d$ is the $d-$sphere in the standard embedding in $\R^{d+1}$ whilst $S^d$ is the family of systems with pure states given by $S^d$ (hence embeddings of $S^d$ in $\R^k$ where $k$ not necessarily equal to $d+1$). This map does not capture all the relations, namely there are `coincidences' like the qubit being both in ${\rm QT}_\C$ and $V_d$.}\label{fig:classification}
\end{figure}

In this work we have introduced new families of systems (see Sections~\ref{sec:twopoint} and~\ref{sec:grassmann}) which generalise previously known systems. In Figure~\ref{fig:classification} we map out the space of transitive systems with compact pure states including the new families of systems introduced in this work. 

The advantage of the methods introduced in this work are two-fold: firstly we can generate examples of non-classical systems and secondly we can systematically classify non-classical systems, thus providing us with a fuller picture of non-classical systems lying beyond quantum theory.

\subsection{The search for alternative theories and the issue of composition}

The tools presented in this work allow us to systematically search for non-classical systems. However it is not certain that these systems compose in a non-trivial manner (existence of entangled states and measurements). For example it is shown in~\cite{masanes_measurement_2019} that the only full theory with systems having the same dynamical structure as quantum theory is quantum theory itself. In~\cite{Masanes_entanglement_2014} it is shown (under certain additional assumptions such as local tomography) that the only systems corresponding to  $d$-balls which compose non-trivially are for $d = 3$. However, removing the requirement of local tomography allows for d-balls to compose non-trivially in at least some cases, including those of real and quaternionic bits~\cite{Barnum_composites_2020} (where a category in which arbitrary d-balls combine via a monoidal product is also described, though the acceptability of these composites is less clear). Out of the family of systems classified in section~\ref{sec:grassmann} it is known that one of them (quartic quantum theory) does not compose under the assumption of local tomography and the requirement that the theory remain quartic~\cite{Zyczkowski_quartic_2008,Lee_higher_2017}. The question remains open as to whether any of the systems with pure states given by Grassmann manifolds compose non-trivially, with or without the assumption of local tomography

\section{Conclusion}~\label{sec:conclusion}

In this work we have introduced the \emph{OPF framework} which is used to characterise systems in GPTs. By separating the dynamical and probabilistic components of systems this framework provides new insight into non-classical systems. It allows us to consider \emph{families} of systems which share a common dynamical structure. We introduce the notion of a \emph{rigid} dynamical structure and show that for such structures one can classify all probabilistic structures using representations of the dynamical group. A key feature of rigid dynamical structures is that they do not admit continuous deformation of probabilistic structure. 

Moreover we introduced multiple new families of non-classical systems, such as the \emph{complex Grassmann} systems. Many of these families contain known non-classical systems, as well as providing infinitely many examples of non-classical systems which were not known. As well as exploring the space of non-classical systems by finding new examples, we mapped out this space in a more systematic manner by introducing families of systems which share a dynamical structure. 

The present work has limited itself to single systems. In general it is not a given that these systems can be made to compose in a non-trivial way (i.e. with entangled states). Future work will involve extending the OPF framework to include composition (along the lines of~\cite{galley_modification_2018,masanes_measurement_2019}) and determining whether any of the new families of systems presented in this work can compose in a non-trivial manner. Finding such examples would provide new non-classical theories whose informational properties could be characterised and contrasted to quantum theory. Alternatively a proof that none of the Grassmann systems can compose would lend credence to the proposition that there are few fully compositional general probabilistic theories.

\section{Acknowledgements}

We thank the referees for their detailed reports which helped substantially improve the present paper, both in terms of fixing errors in proofs and leading to a more rigorous notion of deformation of probabilistic structure. TG acknowledges helpful discussions with Jon Yard.  LM acknowledges financial support by the UK's Engineering and Physical Sciences Research Council [grant number EP/R012393/1]. This research was supported in part by Perimeter Institute for Theoretical Physics. Research at Perimeter Institute is supported by the Government of Canada through the Department of Innovation, Science and Economic Development and by the Province of Ontario through the Ministry of Research, Innovation and Science.

\nocite{apsrev42Control}
\bibliographystyle{apsrev4-2.bst}
\bibliography{refs}

\appendix

\newpage

\section{Notation}\label{app:notation}

\begin{center}
	
	\begin{tabular}{l l}

		$\S_X$ & System with pure states $X$ \\ 
			
		$\Sym(X)$ &  Symmetric group on set $X$ \\ 
		
		$\Diff(X)$ & Group of diffeomorphisms on manifold $X$ \\ 
		
		$\Omega_x$ & Image of $\Omega$ map for en element $x$ \\
		
		$\Omega_X$ & The set $\Omega_x, \forall x \in X$ \\
		
		$\conv\left(\Omega_X\right)$ & Convex hull of  $\Omega_X$  \\
		
		$\{(p_i,x_i)\}_i$ &  Ensemble of pure states \\ 
				
		$\{(p_i,x_i)\}_i \mapsto \sum_i p_i \Omega_{x_i}$ &  Mixed state representation of an ensemble \\
		
		$\F_X$ & set of OPFs (outcome probability functions) for system with pure states $X$ \\ 
		
		$\R[\F_X]$ & $\R[\F_X]: = \{\tilde \f(x)| \tilde \f(x) = \alpha \f_1(x) + \beta \f_2(x), \ \alpha, \beta \in \R, \ \f_1, \f_2 \in F\}$ \\
		
		$\f \mapsto \Lambda_\f$ & Extension of an OPF to ensembles/mixed states \\ 
		
		$(\Gamma,V,\bF)$ & Representation $\Gamma: G \to \GL(V)$ of a group $G$ on linear space $V$ over $\bF$ \\ 
				
		$m(V,V_i)$ & Multiplicity of the sub-representation $V_i$ in the representation $V$ \\
		
		$\Hom_G(V,W)$ & Linear space of $G$-equivariant maps from $V$ to $W$ \\
		
		$W_\C$ & Complexification of a real representation $W$ \\
		
		$V_\R$ & Restriction to reals of a complex representation $V$ \\
		
		$C(X, \bF)$ & Continuous functions $X \to \bF$ for a topological space $X$ and a field $\bF$ \\
			
		$\bF[G/H]$ & Vector space of functions $G/H \to \bF$ for $G,H$ finite \\ 
			
		$V_G$ & G-module (carrier space of a representation of $G$) \\
		
		$\I_V$ & Identity operator on $V$ \\
		
		$\bar V$ & Complex conjugate vector space \\
		
		$\bar v$ & Complex conjugate of a vector v
		
	\end{tabular}

\end{center}

\section{Background group theory and group representation theory}\label{app:group_rep}

We briefly outline a few concepts from group representation theory which will be needed for the proofs. 

\subsection{Representation theory basics}

\begin{definition}[Group representation]
A representation of a group $G$ is a homomorphism $\rho: G \to \GL(V)$, where $V$ is a finite dimensional vector space. When $G$ is a Lie group this is a continuous map. $V$ is called the carrier space of the representation, or a $G$-module.
\end{definition}

Let us denote by $\Hom(V,W)$ the space of linear maps $V \to W$.

\begin{definition}[Intertwining operators]
Let $(\rho,V)$ and $(\sigma,W)$ be representations of $G$. An element $T \in \Hom(V,W)$ is called an intertwining operator if $T \circ \rho = \sigma \circ T$. We denote the linear space of all such maps $\Hom_G(V,W)$.
\end{definition}

$\Hom_G(V,W)$ is a subspace of $\Hom(V,W)$.

\begin{theorem}[Schur's Lemma] 
Let $(\rho,V, \C)$ and $(\sigma,W, \C)$ be irreducible representations of $G$ over $\C$. Then $\dim \ \Hom_G(V,W) = 1$ when $\rho \cong \sigma$ and $\dim \ \Hom_G(V,W) = 0$ otherwise.
\end{theorem}

This implies that $\Hom_G(V,V) = \C \I_V$ for irreducible $V$. Schur's Lemma also has important consequences for reducible representations. For instance consider the case where  $\dim (\Hom_G(V,W)) = 1$ where $V$ is irreducible and $W$ is reducible. Then Schur's Lemma entails that $W$ must contain the irreducible representation $V$ exactly once, and that $\Hom_G(V,W) = \C \I_V$.

\subsection{Left regular and $C(G)$ representations}

\begin{definition}[Left regular representation (finite group)]
	The left regular representation of a finite group $G$ is given by:
	\begin{align}
	& \rho: G \to \GL(\C[G]) \ , \\
	& \rho_g \ket{g'} =	  \ket{gg'} \ ,
	\end{align}
	where $\C[G]$ is a complex linear space spanned with orthonormal basis $\{\ket g\}_{g \in G}$.
\end{definition}

Here the action of $G$ on $\C[G]$ is just permutation of the basis vectors.

\begin{definition}[Left regular representation (compact group)]
	The left regular representation of a compact topological group $G$ is given by:
	\begin{align}
     \rho_g f(g') =	 f(g^{-1} g') \ ,
	\end{align}
	where $f \in C(G, \C)$.
\end{definition}

$\rho$ is a continuous homomorphism: $\rho_g \rho_h f(g') = \rho_h f(g^{-1}g') = f(h^{-1}g^{-1} g') = f((gh)^{-1} g') = \rho_{gh} f(g')$.

\subsection{Restricted and induced representations}

\subsubsection{Restricted representation}

\begin{definition}[Restricted representation]
	The restriction of a representation $(\rho,V)$ of a group $G$ to a subgroup $H$ is a representation of $H$ on $V$ with operators $\rho(h): V \to V$ for all $h \in H$. It is denoted $\rho_{|H}$.
\end{definition}

In general the restriction of an irreducible representation $G$ will give a reducible representation $H$.

\begin{example}
	Consider the fundamental representation of $\SO(3)$ on $\R^3$ and restrict to the subgroup $\SO(2)$ with matrices:
	\begin{align}
	\begin{pmatrix}
	1 & 0 & 0 \\
	0 & \cos(\theta) & -\sin(\theta) \\
	0 & \sin(\theta) & \cos(\theta)
	\end{pmatrix} \ .
	\end{align}
	This is a reducible representation of $\SO(2)$ containing the trivial and fundamental representation.
\end{example}

\subsubsection{Induced representation}

\begin{definition}[Induced representation: $G$ and $H$ finite]
	A representation $\rho: H \to \GL(V)$ of a subgroup $H$ of $G$ induces a representation of $G$ given by the right regular representation of $G$ on:
	\begin{align}
	\Ind_H^G(V) = \{\phi: G \to V|\phi(hg) = \rho(h) \phi(g), \forall h \in H, g \in G   \} \  .
	\end{align}	
\end{definition}

The carrier space $\Ind_H^G(V)$ is the space of functions $G \to V$ which are $H$-equivariant. We can write it as $\Ind_H^G(V)  \cong \Hom_H(G,V)$

\begin{definition}[Induced representation: $G$ locally compact and $H$ closed]
	A representation $\rho: H \to \GL(V)$ of a subgroup $H$ of $G$ induces a representation $(\sigma,\Ind_H^G(V))$ of $G$:
	\begin{align}
	\Ind_H^G(V) = \{\phi: G \to V|\phi(gh^{-1}) = \sigma(h) \phi(g), \forall h \in H, g \in G , \ \phi \ {\rm continuous}    \} \ .
	\end{align}
\end{definition}

The action of $G$ on $\Ind_H^G(V)$ is given by $ \sigma(g) \phi(g') = \phi(g^{-1}g')$ where $\sigma: G \to \GL(\Ind_H^G(V))$.

\subsection{Frobenius reciprocity}

\begin{theorem}[Frobenius reciprocity]\label{thm:Frob_rec}~\cite[Theorem 7.47]{sepanski_compact_2007}
	Take $G$ a finite group and $H$ a subgroup or G a Lie group and $H$ a closed subgroup of $G$. Given $V$ a representation of $H$ and $W$ a representation of $G$, then as vector spaces
	\begin{align}
	\Hom_G(W, \Ind_H^G(V)) \cong \Hom_H(W_{|H}, V) \ .
	\end{align}
\end{theorem}

We define $\C[G/H]$ (with $G,H$ finite) to be the space of functions $G/H \to \C$ which forms a vector space under pointwise addition and scalar multiplication. It carries a representation of $G$, inherited from the action of $G$ on $G/H$. For $G/H$ a topological space, $C(G/H, \C)$ is the vector space of continuous functions $G/H \to \C$, also carrying a representation of $G$.

\begin{corollary}~\label{cor:mult_in_c_decomp}
	For $(G,H)$ finite the irreducible representation $W$ occurs in $\C[G/H]$ with multiplicity equal to the multiplicity of the trivial representation in $W_{|H}$. For $(G,H)$ compact the irreducible representation $W$ occurs in $C(G/H, \C)$ with multiplicity equal to the multiplicity of the trivial representation in $W_{|H}$.
\end{corollary}

\begin{proof}
	First observe for $(G,H)$ finite $\Ind_H^G(\C) = \C[G/H]$ and for $(G,H)$ compact $\Ind_H^G(\C) = C(G/H, \C)$.
	
	Let us consider the case where $V$ is the trivial representation: $V \cong \C$. Then $W_{|H}$ contains a trivial representation with multiplicity $m(W_{|H},1_H)$. Using Schur's lemma the only $H$-equivariant map from $W_{|H}$ to $V$ are identities on the trivial representations. In other words $\Hom_H(W_{|H}, \C) = \C^{m(W_{|H},1_H)}$.
	
	Similarly we have that $\Hom_G(W,\Ind_H^G(\C)) =\C^{m(\Ind_H^G(\C), W)}$ where $m(\Ind_H^G(\C), W)$ is the multiplicity of $W$ in $\Ind_H^G(\C)$. By  Theorem~\ref{thm:Frob_rec} we obtain:	
	\begin{align}
	m(\Ind_H^G(\C), W) = m(W_{|H},1) \ .
	\end{align}
\end{proof}

\section{Representations over $\R$}\label{app:real_rep}

A representation $\Gamma$ is a homomorphism $G \to \GL(V)$ for some vector space $V$. Typically $V$ is assumed to be a vector space over $\C$, and many tools in representation theory apply to this case. Schur's lemma for instance holds for representations over the complex field, but not always for those over $\R$. For a Gelfand pair $(G,H)$ the property of having a trivial irreducible representation when restricted to $H$ which is of multiplicity $0$ or $1$ holds for representations over $\C$, and not necessarily $\R$. 

In this section we explore some of the subtleties involved in dealing with representations over $\R$ and prove some lemmas which will be needed for Theorem~\ref{thm:classification}. First we present an example to introduce some of the relevant concepts.

\begin{example}[Fundamental representation of $\SO(2)$]

Consider the representation of $\SO(2)$ over $\R^2$:

\begin{align}
\Gamma(\theta) = 
\begin{pmatrix}
\cos (\theta) & - \sin (\theta) \\
\sin (\theta) & \cos(\theta)
\end{pmatrix} \ . 
\end{align}

This representation is irreducible over $\R^2$. However consider this representation acting on $\C^2$ (obtained from $\R^2$ by allowing complex linear combinations of the basis elements). Then there exist the following matrices $S$ and $S^{-1}$:

\begin{align}
S = 
\begin{pmatrix}
-i & i \\
1 & 1
\end{pmatrix} \ , \ 
S^{-1} = 
\begin{pmatrix}
\frac{i}{2} & \frac{1}{2}  \\
-\frac{i}{2}  & \frac{1}{2}
\end{pmatrix} \ ,
\end{align}
such that $\Gamma'(\theta) = S^{-1} \Gamma(\theta) S$, with 
\begin{align}
\Gamma'(\theta) = 
\begin{pmatrix}
e^{-i \theta} & 0 \\
0 & e^{i \theta}
\end{pmatrix} \ .
\end{align}
So the irreducible representation over $\R^2$ is reducible over $\C^2$. Consider once again the irreducible representation $\Gamma$ over $\R$. Then this commutes with all matrices proportional to the identity, but also matrices proportional to $J$, where $J$ is:
\begin{align}
J = 
\begin{pmatrix}
0 & 1 \\
-1 & 0
\end{pmatrix} \ .
\end{align}
Moreover one can show that only matrices which are linear combinations of $J$ and $\I$ commute with the whole group.
\end{example}

\subsection{Definitions}

\begin{definition}[Real, complex and quaternionic structure]
	Consider an irreducible representation $\rho,V$ with $V$ a complex vector space. Then $V$ has a real structure if there exists an equivariant anti-linear map $j: V \to V$ such that $j^2 = \I$, $V$ has a quaternionic structure if there exists an equivariant anti-linear map $j: V \to V$ such that $j^2 = -\I$. Otherwise $V$ has a complex structure.
\end{definition}

\begin{lemma}[Descent map]
	Given a representation $(\Gamma,V,\C)$ of a group $G$ over a complex space $V$ equipped with a real structure $j$ the projection $V \to V^j$ with kernel the $j = -1$ eigenspace of $V$ is a descent map.  $V^j: \{v \in V: j(v) = v\}$ carries a real representation of $G$.
\end{lemma}

\begin{proof}

	Consider $V^j: \{v \in V: j(v) = v\}$. Observe that this set is closed under real linear combinations, and not complex linear combinations. As such it has the structure of a real vector space. Moreover $V^j \otimes_\R \C \cong V$. 
	
Since $j(gv) = gj(v)$, $V^j$ carries an action of $G$. Moreover for $v_1, v_2 \in V^j$ we have $a g v_1 + b g v_2 = a gj(v_1) + b gj(v_2) = g j(a v_1 + b v_2) = g(a v_1 + b v_2)$ for $a ,b \in \R$. Therefore $V^j$ carries a real representation of $G$.

\end{proof}

Typically $j$ is the conjugation map: $j(v) = \bar v$. We see that it is equivariant when $j(gv) = \bar g \bar v$, i.e. the representation $g$ and $\bar g$ are isomorphic.

\begin{definition}[Complexification]
	A real vector space $V$ has complexification $V_\C := V \otimes \C$ where $V_\C$ is a complex vector space. A basis for this complex vector space is $v \otimes 1$. $\dim_\R(V) = \dim_\C(V_\C)$.
\end{definition}

We sometimes write $V_\C \cong V \oplus i V$.

\begin{definition}[Restriction of scalars]
	A complex vector space $V$ is isomorphic (as a real vector space) to the real vector space $V_\R$. $\dim_\R(V_\R) = 2 \  \dim_\C(V)$. If $\{e_i\}$ is a basis for $V$, then $\{\{e_i\},\{i e_j\}\}$ is a basis for $V_\R$.
\end{definition}

Consider $V$ a complex vector space and $W$ a real vector space. 

\begin{align}
(V_\R)_\C \cong_\C V \oplus \bar V \ , \\
(W_\C)_\R \cong_\R W \oplus W \ ,
\end{align}
where $\bar V$ is the complex conjugate vector space of $V$. $\bar V$ has the same elements and addition rule as $V$, but scalar multiplication is given by $\lambda \cdot v = \bar \lambda V$.

\begin{theorem}\cite[Theorem 3.37, p.41]{Fulton_representation_1991}\label{thm:real-comp-quat}
	For an irreducible representation  $(\rho,W, \R)$ the action of $G$ extends naturally to $W_\C$. It has one of the following decompositions:
	\begin{enumerate}
		\item $W_\C  \cong V$ for a complex irreducible representation $V$ if and only if $V$ has a real structure.
		\item $W_\C  \cong V \oplus \bar V$ for a complex irreducible representation $V$ if and only if $V$ has a complex structure.
		\item $W_\C  \cong V \oplus V$ for a complex irreducible representation $V$ if and only if $V$ has a quaternionic structure.
	\end{enumerate}
\end{theorem}

Every real irreducible representation $(\rho,W, \R)$  can be obtained from irreducible complex representations of one of the above three forms, by taking the descent map $W_\C^j$.

\begin{corollary}
	Consider the set of all irreducible complex representations of a group $G$, these are the representations $\{V_i^r\}$ of real type, the representation  $\{V_j^c\}$ complex type and the representations $\{V_j^q\}$ of quaternionic type. The set of all real irreducible representations has a unique representative element in $\{V_i^r\} \bigcup \{V_i^c/*\} \bigcup \{V_j^q\}$, where $ \{V_i^c/*\}$ is the set of complex representations where conjugate representations are equivalent. 
\end{corollary}

\subsection{Lemmas}

In the following section $G$ is assumed finite or compact. 

\begin{lemma}\label{lem:rea_comp_struc}
	For a Gelfand pair $(G,H)$ all irreducible representations $(\Gamma,V, \C)$ with an $H$-invariant subspace  have either a real or complex structure.
\end{lemma}

\begin{proof}
	For $(G,H)$ finite every representation $(\Gamma,V, \C)$  with an $H$-invariant subspace occurs exactly once in the decomposition of $\C[G/H]$ by Corollary~\ref{cor:mult_in_c_decomp}.  $\C[G/H]$ has a real structure (compatible with the group action) and decomposes as $\R[G/H] \oplus i \R[G/H] \cong \R[G/H]_\C$.

	By Theorem~\ref{thm:real-comp-quat} the complexification map $\R[G/H] \to \R[G/H]_\C \cong \C[G/H]$ sends every real irreducible representation $W$  in $\R[G/H]$ to either:
	
	\begin{enumerate}
		\item An irreducible representation $W_\C \cong V$ in $\C[G/H]$ if $W$ is real.
		\item A reducible representation $W_\C \cong  V \oplus \bar V$ in $\C[G/H]$ if $W$ is complex.
		\item A reducible representation $W_\C \cong  V \oplus V$ in $\C[G/H]$ if $W$ is quaternionic.
	\end{enumerate}
	However since $\C[G/H]$ only contains single copies of irreps it cannot contain  $V \oplus V$ and hence no irrep $W$ can be of quaternionic form. 
	
	For $(G,H)$ compact every representation $(\Gamma,V, \C)$  with an $H$-invariant subspace occurs exactly once in the decomposition of $C(G/H, \C)$ by Corollary~\ref{cor:mult_in_c_decomp}. $C(G/H, \C)$ is the complexification of $C(G/H, \R)$ and the same proof as for the finite case follows through.
\end{proof}

\begin{lemma}\label{lem:Hinv_real_sub}
	For $(G,H)$ a Gelfand pair and $(\Gamma,V, \R)$ an irreducible representation over $\R$, then the following holds:
	\begin{enumerate}
		\item If $\Gamma$ is irreducible over $V_\C$ then the (real) dimension of the $H$-invariant subspace in $V$ is the same as the (complex) dimension of the $H$-invariant subspace in $V_\C$: $\dim_\R(\Hom_H(\Gamma, 1_H) = \dim_\C(\Hom_H(\Gamma, 1_H)$. It is either $0$ or $1$.
		\item If $\Gamma$ is reducible over $V_\C$ then the (real) dimension of the $H$-invariant subspace in $V$ is twice that of the (complex) dimension of the $H$-invariant subspace in $V_\C$: $\dim_\R(\Hom_H(\Gamma, 1_H) =2 \ \dim_\C(\Hom_H(\Gamma, 1_H)$. It is either $0$ or $2$. 
	\end{enumerate}
\end{lemma}

\begin{proof}
	\begin{enumerate}
	\item  $V_\C = V \otimes_\R \C$. The $H$-invariant complex subspace $V_H'$ of $V_\C$ is 0 or 1 dimensional. Consider the 1 dimensional case, then $V_H'$ is spanned by a vector $v_H' = v_H \otimes_\R \gamma$. Now since $V \cong V_\C^j$ for the map $j(v') = \bar v'$, we have that under the descent map the subspace $V_H'$ is mapped to the (real) one dimensional subspace $V_H$ of $V$, spanned by $v_H \otimes_\R \gamma$ with $\gamma \in \R$. Any vector $v$ such that $j(v') = v'$ which is not in $V_H'$ will be mapped to a vector in $V$ which is not $H$-invariant (since $j$ is equivariant). As such the subspace of all $H$-invariant vectors in $V$ is just $V_H$ and is one dimensional. The case where $V_H'$ is of dimension $0$ is immediate.

	\item 	By Lemma~\ref{lem:rea_comp_struc} if $\Gamma$ is irreducible over $\R$ but reducible over $\C$ then it has a complex structure and is of the form $\rho \oplus \rho^*$ where $\rho$ and $\rho^*$ are irreducible. Since $(G,H)$ is a Gelfand pair either both $\rho$ and $\rho^*$ contain a one dimensional $H$-invariant complex subspace or neither do. Let us consider the case where they both do.  
	\begin{align}
	\Gamma'= 
	\begin{pmatrix}
	\rho & 0 \\
	0 & \rho^*
	\end{pmatrix} \ .
	\end{align}
	We use a matrix representation where the entries of $\rho^*$ are the complex conjugates of those of $\rho$. 
	Let $v_i$ be a basis for $V$, then 
	\begin{align}
	v_i^+ = \begin{pmatrix}
	v_i \\
	\bar v_i
	\end{pmatrix} \ ,
	v_i^- =  \begin{pmatrix}
	v_i \\
	- \bar v_i
	\end{pmatrix} \ ,
	\end{align}	
	form a basis for $V \oplus \bar V$. By also considering the elements:
	\begin{align}
	i v_i^+ = 	\begin{pmatrix}
	i v_i \\
	i \bar v_i
	\end{pmatrix} \ ,
	i v_i^- =
	\begin{pmatrix}
	i v_i \\
	- i \bar v_i
	\end{pmatrix} \ ,
	\end{align}
	we obtain a real basis for $V \oplus \bar V$, i.e. any $w \in V \oplus \bar V$ is a real linear combination of 	the above vectors. 
	
	Let
	\begin{align}
	\begin{pmatrix}
	v \\
	\bf 0
	\end{pmatrix} \ ,
	\end{align}
    be an $H$ invariant vector, then so is:
    \begin{align}
    \begin{pmatrix}
	\bf 0 \\
	\bar v
	\end{pmatrix} \ .
    \end{align}
	
	All $H$-invariant vectors are of the form:
	\begin{align}
	v_H = 
	\begin{pmatrix}
	\alpha v \\
	\bar \beta \bar v
	\end{pmatrix} \ ,
	\end{align}	
	
	since the $H$ invariant subspaces in $V$ and $\bar V$ are one dimensional.

	Now we consider the change of basis given by matrices $S$ and $S^{-1}$:
	\begin{align}
	S = 
	\begin{pmatrix}
	-i \I & i \I \\
	\I  & \I 
	\end{pmatrix} \ , \
	S^{-1} = 
	\begin{pmatrix}
	\frac{i \I}{2} & \frac{\I}{2}  \\
	-\frac{i \I}{2}  & \frac{\I}{2}
	\end{pmatrix} \ ,
	\end{align}
	and apply $ S \Gamma' S^{-1}$ to obtain 
	\begin{align}
	\Gamma = 
	\begin{pmatrix}
	\frac{\rho + \rho^*}{2} & \frac{i (\rho^* - \rho)}{2} \\
	\frac{i(\rho - \rho^*)}{2} & \frac{\rho + \rho^*}{2}
	\end{pmatrix} \ ,
	\end{align}
	has real valued entries. The action of $S$ on the basis is:
	\begin{align}
	S v_i^+ =  w_i^+ = 
	\begin{pmatrix}
	 i (\bar v_i - v_i)\\
	v_i + \bar v_i
	\end{pmatrix}  \quad
	S v_i^- =  w_i^- = 
	\begin{pmatrix}
	- i (v_i  + i \bar v_i)\\
	v_i - \bar v_i
	\end{pmatrix}  \ ,
	\end{align}	
	Considering the real basis elements we also obtain:
	\begin{align}
	iw_i^+ = 
	\begin{pmatrix}
	 v_i  -  \bar v_i\\
	i (v_i +\bar v_i)
	\end{pmatrix}  \ ,
	iw_i^- = \begin{pmatrix}
	 v_i  + \bar v_i\\
	i (v_i - \bar v_i)
	\end{pmatrix}  \ .
	\end{align}
	Under the real structure $j(v) = \bar v$ the basis vectors $w_i^+$ and $i w_i^-$ are $+1$ eigenvectors of $j$, and form a basis for $V^j \cong \R^n$. The other basis vectors are $-1$ eigenvectors and form a basis for $i \R$. Since $\Gamma = \bar \Gamma$ we have that $j(v)=v \implies j(gv)=gv$ and so the subspace $V^j$ is closed under $G$.

	The image of the $H$-invariant vectors under $S$ is:
	\begin{align}
	S v_H(\alpha,\beta) = w_H(\alpha,\beta) =
	\begin{pmatrix}
	- i \alpha v + i \bar \beta  \bar v \\
	\alpha v + \bar \beta \bar v 
	\end{pmatrix} \ .
	\end{align}
	These are invariant under $j$ for $\beta = \alpha$. Any such $j$ invariant vector is a real linear combination of  $w_H^1 = w_H(\alpha = 1,\beta=1)$ and  $w_H^2 = w_H(\alpha  = i, \beta = i)$. As such the $H$-invariant subspace in $V^j$ is two dimensional with basis vectors written out explicitly as:
	\begin{align}\label{eq:inv_basis_vec}
	w_H^1 = 
	\begin{pmatrix}
	i (\bar v - v) \\
	v + \bar v
	\end{pmatrix} \ , \\
	w_H^2 = 
	\begin{pmatrix}
	v + \bar v \\
	i ( v - \bar v)
	\end{pmatrix} \ .
	\end{align}
\end{enumerate}
\end{proof}

\begin{corollary}\label{cor:commute-H_vec}
	For $(G,H)$ a Gelfand pair and $\Gamma$ an irreducible representation over $\R^n$, then all $H$-invariant vectors are related by an invertible transformation which commutes with $\Gamma$.
\end{corollary}

\begin{proof}
In the case where $\Gamma$ is complex irreducible this is immediate from Schur's lemma. In the case where $\Gamma$ is complex reducible, then its complexification is reducible:
	\begin{align}
	\Gamma'= 
	\begin{pmatrix}
	\rho & 0 \\
	0 & \rho^*
	\end{pmatrix} \ .
	\end{align}
	By Schur's lemma all matrices which commute with the whole group are of the form:
	\begin{align}
	M'(\alpha , \beta)= 
	\begin{pmatrix}
	\alpha \I  & 0 \\
	0 & \beta \I
	\end{pmatrix} \ .
	\end{align}

	Using the transformation $S$ of the previous proof one obtains: 
	\begin{align}
	M(\alpha,\beta) = SM'(\alpha,\beta) S^{-1} = 
	\begin{pmatrix}
	\frac{\alpha + \beta}{2} \I & \frac{i (\beta - \alpha)}{2} \I \\
	\frac{i(\alpha -\beta)}{2} \I & \frac{\alpha + \beta}{2} \I
	\end{pmatrix} \ ,
	\end{align}
	which all commute with $\Gamma$. This is real valued for $\alpha = \beta=1$, $\alpha = -\beta = i$ and all real linear combinations of these matrices. $M(1,1) = \I$ and $M(i,-i) = J$ where 
	\begin{align}
	J = 
	\begin{pmatrix}
	0 & -1 \\
	1 & 0 
	\end{pmatrix}
	\end{align}
	Consider the two basis $H$ invariant vectors $w_H^1$ and $w_H^2$ of Equation~\eqref{eq:inv_basis_vec}.	These are related by $-M(i,-i)$. Therefore all linear combinations of these are related by real linear combinations of $M(1,1)$ and $M(i,-i)$.	
\end{proof}

\begin{corollary}
	For $\Gamma, W$ an irreducible representation over $\R^n$, then the following holds:
	\begin{enumerate}
		\item If $W_\C$ is irreducible then  $\Hom_G(\R^n,\R^n) = \R$ . The only equivariant homomorphisms are scalar multiples of the identity.
		\item If $W_\C$ is reducible into irreducible representations with complex structure then $\Hom_G(\R^n,\R^n) \cong \R^2$. 
	\end{enumerate}
\end{corollary}

\begin{lemma}\label{lem:gelfand_decomp_finite}
	For a Gelfand pair of finite groups $(G,H)$, the representation of $G$ acting on $\R[G/H]$ contains exactly once every irreducible representation of $G$ which has a trivial sub-representation when restricted to $H$ and does not contain any other representations.
\end{lemma}

\begin{proof}
By Corollary~\ref{cor:mult_in_c_decomp}  $\C[G/H]$ contains every irreducible representation $W$ with an $H$ invariant subspace with multiplicity 1, if an irreducible is of complex type then its complex conjugate is necessarily also in $\C[G/H]$, since if $W$ has a $H$-invariant vector so does $W^*$.

We have $\C[G/H] = \R[G/H]_\C$ which contains only irreducible representations with complex or real structures by Lemma~\ref{lem:rea_comp_struc}. Therefore we have:
\begin{align}
\C[G/H] = \bigoplus V_i \oplus \bigoplus (U_j \oplus \bar U_j) \ ,
\end{align}
where $V_j$ are the irreducible representations of real type, and $U_j$ are irreducible representations of complex type. There are no degeneracies.

$\R[G/H]$ decomposes into real irreducible representations $W_i$.
\begin{align}
\R[G/H] & = \bigoplus_i W_i \ , \\ 
\C[G/H]  = \R[G/H]_\C & = \R[G/H]  \otimes \C = \bigoplus_i (W_i \otimes \C) \ .
\end{align}
Hence every irreducible representation $W_i$ is sent to $(W_i)_\C$ in $\C[G/H]$. By Theorem~\ref{thm:real-comp-quat} these are of the form $V$ for $V$ irreducible of real type, $V \oplus \bar V$ for $V$ of complex type and $V \oplus V$ for $V$ of quaternionic type, where we know by Lemma~\ref{lem:rea_comp_struc} that this latter case does not occur.

Combining the above:
\begin{align}
\bigoplus_i (W_i \otimes \C) \cong \bigoplus V_j \oplus \bigoplus (U_k \oplus \bar U_k) \ ,
\end{align}
where every real irreducible subspace $(W_i \otimes \C)$ corresponds to either a subspace $V_j$ or  $(U_k \oplus \bar U_k)$. Since these occur with multiplicity 1, every $(W_i \otimes \C)$ occurs with multiplicity one in $\R[G/H]_\C$. This implies that every $ W_i$ occurs with multplicity 1 in $\R[G/H]$.
\end{proof}

\begin{lemma}\label{lem:gelfand_decomp_cont}
	For a Gelfand pair $(G,H)$, where $G$ compact and $H$ a closed subgroup, the representation of $G$ acting on $C(G/H,\R)$ contains exactly once every irreducible representation of $G$ which has a trivial sub-representation when restricted to $H$ and does not contain any other representations.
\end{lemma}

\begin{proof}
Direct from proof of Lemma~\ref{lem:gelfand_decomp_finite} replacing $\C[G/H]$  with $C(G/H,\C)$ and $\R[G/H]$ with  $C(G/H,\R)$
\end{proof}

\section{Convex linear state space}\label{app:statespace}

In this appendix we derive the convex linear state space associated to a system $\S_X = \{X,G,\F_X\}$.

\begin{definition}[Preparation space]
	The preparation space of $\S_X = \{X,G,\F_X\}$ consists of the space of all ensembles $\{(p_i, x_i)\}_i$ for $X$ finite, which is affinely equivalent to the simplex over the set $X$ (i.e. with the simplex with $|X|$ extremal points, corresponding to the Dirac measures on $X$). When $X$ is a topological space, the preparation space consists of all positive normalised measures on $X$, which is the generalised simplex $M_1^+(X)$ 
\end{definition}

In some treatments~\cite{Holevo_probabilistic_2011} a mixture space is considered, where the mixing operation applies also to ensembles. For example one can take $j$ ensembles $\{(p_i^j, x_i^j)\}_i$ to prepare an ensemble $\{(p_j,\{(p_i^j, x_i^j)\}_i)\}_j$. In our case we can consider ensembles of ensembles by setting $\{(p_j,\{(p_i^j, x_i^j)\}_i)\}_j = \{(p_j p_i^j,x_i^j)\}_{ij}$ which is an ensemble of pure states and hence in the preparation space. We note that in a mixture space $\{(p_j,\{(p_i^j, x_i^j)\}_i)\}_j$ and $\{(p_j p_i^j,x_i^j)\}_{ij}$ are two distinct elements, even if there can be no affine functional which separates them, by definition. In the ``maximally separated" classical state space, these two ensembles are always equivalent. Therefore there is no loss of generality when considering a preparation space, as opposed to a mixture space.

\subsection{Proof of Theorem~\ref{thm:sys_maps}}

\begin{proof}

We consider a system  $\S_X = \{X,G,\F_X\}$  where $\R[\F_X]$ is finite dimensional as a linear space. We consider only a preparation space of finite ensembles for this proof (one can extend to continuous ensembles, but for finite dimensional systems this is not necessary). One can extend the OPFs to the preparation space:
\begin{align}
P(\f|\{(p_i , x_i)\}_i) = \sum_i p_i \f(x_i) \ .
\end{align}
Two ensembles $\{(p_i , x_i)\}_i$ and $\{(p_j' , x_j')\}_j$ where $P(\f|\{(p_i , x_i)\}_i) = P(\f|\{(p_j' , x_j')\}_j)$ for all $\f \in \F$  are equivalent ensembles. We can define the linear form $\Omega_x \in \R[\F_X]^*$: $ \Omega_x(\f) = \f(x) \ \forall \f \in \F_X$. 
\begin{align}
P(\f|\{(p_i, x_i)\}_i) = \sum_i p_i \f(x_i) = (\sum_i p_i  \Omega_{x_i})(\f) \ .
\end{align}
Hence to the ensemble  $\{(p_i, x_i)\}_i$ we associate the mixed state $\omega = \sum_i p_i \Omega_{x_i}$, which is naturally an element of $\R[\F_X]^*$. Any two indistinguishable ensembles are mapped to the same mixed state.  Now we can map the OPFs to linear functionals $\Lambda_\f$ in $(\R[\F_X]^*)^*$:
\begin{align}
\Lambda_\f(\omega) = P(\f|\omega) ,\ \forall \omega \in \conv(\Omega_X) \ .
\end{align}
By assumption $\Omega_x \to \Omega_{gx}$ is a continuous group action when $G$ is a continuous group. Moreover it extends to $\conv(\Omega_X)$ as follows
\begin{equation}
\sum_i p_i \Omega_{x_i} \xrightarrow{g} \sum_i p_i \Omega_{g x_i} \ .
\end{equation}
This uniquely extends to $\sp(\Omega_X) \cong \R[\F_X]^*$ and hence $\Gamma: G \to \GL(\R[\F_X]^*)$ is a group representation.

\end{proof}

\subsection{Proof of Lemma~\ref{lem:extremal_points}}\label{app:extremal_points_lem}

\begin{proof}
Let $\Omega, \Gamma, \Lambda$ be an embedding of $\S_X = \{X, G , \F_X\}$ into $(V,V^*)$ with $V \cong \R^d$.

$\Omega_X$ can be generated by applying $\Gamma_g$ to a reference vector $\Omega_{x_0}$ for all $g \in G$. Since $G$ is compact the matrices $\Gamma_g$ can be made orthogonal. Hence every $\Omega_x = \Gamma_g \Omega_{x_0}$ for some $g \in G$ is such that $\Omega_x^T \Omega_x = \Omega_{x_0}^T \Omega_{x_0}$. All $\Omega_x$ are such that $\u \cdot \Omega_x = 1$ with $\u$ the unit effect, therefore $\Omega_X$ lies in the affine span of $\Omega_X$ which is a $d-1$ dimensional hyperplane. This is the hyperplane composed of all $v \in V$ such that $\u \cdot v = 1$. Every $\Omega_x$ can be written as:
\begin{align}
\begin{pmatrix}
1 \\
\tilde \Omega_x
\end{pmatrix}
\end{align}
in a basis where the first element corresponds to the subspace spanned by the unit effect. Observe that $\Gamma_G$ acts trivially on the subspace spanned by the unit effect, and hence decomposes as $\Gamma = 1 + \tilde \Gamma$ where $1$ is the trivial representation. Observe that the maximally mixed state $\omega = \int_{g \in G} \Gamma_g \Omega_x dg$ for an arbitrary $\Omega_x$ is invariant under $G$. Hence it lies fully in the subspace spanned by the unit effect. Therefore $\omega$ can be written as:
\begin{align}
\begin{pmatrix}
1 \\
\tilde \omega
\end{pmatrix}
\end{align}
where $\tilde \omega$ is the zero vector.

Restricting our attention to the affine span of $\Omega$, i.e. ignoring the unit effect subspace we observe that the matrices $\tilde \Gamma$ are orthogonal, hence $\tilde \Gamma_g \tilde \Omega_{x_0}$ lies on a $d-1$ sphere centred on $\omega$.

Since $\Omega_X$ lies on a hypersphere, this implies that no point in $\Omega_X$ lies inside $\conv(\Omega_X)$, proving the lemma. 

\end{proof}

\subsection{Proof of Lemma~\ref{lem:comp_metric}}\label{app:lem_comp_metric}

\begin{proof}
In the representation in terms of fiducial OPFs the vectors $\conv(\Omega_X)$ are bounded. Moreover they are closed, since any limit of physically realisable extremal preparations is indistinguishable from an extremal physical preparation. Hence $\conv(\Omega_X)$ is a closed and bounded subset of $\R[\F_X]^*$ (isomorphic to $\R^n$ by the assumption of ``Possibility of state estimation using a finite outcome set''), and by the Heine-Borel theorem it is compact in the induced (subspace) topology.

Since $\conv(\Omega_X)$ is bounded  this implies that the absolute value of the matrix entries of $\conv(\Gamma_G)$ are also bounded. Moreover we assume that from a physical point of view the space of transformations $\conv(\Gamma_G) \subset \R^n$ must be topologically closed (in the vector space topology of the space $\R^{n^2}$ of linear transformations $\R^n \to \R^n$), since any mathematical transformation which can be arbitrarily well approximated by physical transformations is indistinguishable from a physical transformation. This implies that the set $\conv(\Gamma_G) \subset \R^{n^2}$ is bounded and closed, and hence compact by the Heine-Borel theorem.
\end{proof}

\subsection{Proof of Theorem~\ref{thm:classification}}\label{app:classification_thm}

\begin{proof}
\begin{enumerate}[wide, labelwidth=!, labelindent=0pt,label = \roman*.]
	\item Take $\S = \{X\cong G/H,\F\}$ with $(G,H)$ a Gelfand pair and $\R[\F]$ finite. To every $x \in X$ we associated a linear functional $\Omega_x \in \R[\F]^*$. Let us call $\Gamma$ the representation of $G$ acting on $\R[\F]^*$. In general $\Gamma$ may be reducible:
	\begin{align}
	\Gamma = \bigoplus_i \Gamma^i \ .
	\end{align}
	We can also decompose states:
	\begin{align}
	\Omega_x = \bigoplus_i \Omega_x^i \ .
	\end{align}
	We have the following equality:
	\begin{align}
	\Gamma_g \Omega_x = \Omega_{gx} \ .
	\end{align}
	Since $x$ is stabilized by $H_x$ (where $H_x$ is stabilizer of $x$: $h x = x$ for all $h \in H_x$) we have 
	\begin{align}
	\Gamma_h \Omega_x = \Omega_x, \ \forall h \in H \ .
	\end{align}
	This implies
	\begin{align}
	\Gamma_h^i \Omega_x^i = \Omega_x^i, \ \forall h \in H \ .
	\end{align}
	This implies that each $\Gamma_{|H}^i$ has at least one trivial sub-representation. 
	\item Let us consider a representation of $G$ acting on a real vector space $V$:
	\begin{align}
	\Gamma = \bigoplus_i \Gamma_i
	\end{align}
	such that each $\Gamma_i$ has at least one $H$-invariant subspace. Take a reference vector $v \in V$ which has support only in the $H$-invariant subspaces, and has support in each subspace $V_i$. By applying $\Gamma_G$ to $v$ we obtain $\Omega_{G/H} \in V$, where we observe that $\conv \left(\Omega_{G/H}\right)$ has $\Omega_{G/H}$ as extremal points, since $\Gamma_G$ can be expressed in orthogonal matrices and hence $\Omega_{G/H} \subset S^n$ (a hyper-sphere in the affine span of the normalised states, centred on the maximally mixed state) as proven in Lemma~\ref{lem:extremal_points}. By taking the convex set of all linear functionals which give values in $[0,1]$ we obtain $\Lambda_\F$ a probabilistic structure. 
	\item Since $(G,H)$ is a Gelfand pair, all real irreducible representations $V_i$  are such that any pair of $H$-invariant vectors are related by an invertible linear transformation $L_i$.  For a representation:
	\begin{align}
	\Gamma = \bigoplus_i \Gamma^i \ ,
	\end{align}
	take two $H$-invariant vectors $v$ and $v'$ which have support in every irreducible subspace. These are related by a transformation $L$: $L v = v'$ which commutes with the group action. We call their orbits under $\Gamma_G$ $\Omega_X$ and $\Omega_X'$. Since $L$ commutes with the group action $L \Omega_X = \Omega_X'$. Let us consider the unrestricted effect spaces for both: $\Lambda_\F$ and $\Lambda_{\F'}'$. $\Lambda_{F'}'(\Omega_X') = \Lambda_{\F'}'(L \Omega_X)$. The set of all effects on $L \Omega_X$ is just $\Lambda_F L^{-1}$, hence $\Lambda_{\F'}' = \Lambda_\F L^{-1}$.
	\item Let us take the case where $(G,H)$ is not a Gelfand pair. There exist (complex) irreducible representations $W$ such that $W_{|H}$ contains more than one trivial sub-representation. 
	
	One can obtain a real irreducible representation from $W$ by one of the three following methods:
	
	\begin{enumerate}
		\item Restriction to $\R$ of $W$ if $W$ is of real type.
		\item Restriction to $\R$ of $W \oplus \bar W$ if $W$ is of complex type.
		\item Restriction to $\R$ of $W \oplus W$ if $W$ if of quaternionic type
	\end{enumerate}
	
	In each case the real irreducible representation $V$ obtained is such that it has invariant $H$-vectors which are not related by a transformation which commutes with the group representation. Let us fix a basis and consider the matrices $\Gamma_g$ acting on $V$. Let us pick two $H$-invariant vectors $v$ and $v'$ such that there is no transformation which commutes with the group action such that $L v = v'$. 

	We now show that $v$ and $v'$ can be used (along with $\Gamma_G$) to define two tomographically inequivalent probabilistic structures, associated with the same carrier space $V$.
	
Let us  call $x_0$ an element in $X$ stabilised by $H$ and define $\Omega_{x_0} = v$ and $\Omega_{x_0}' = v'$. Furthermore let us define $\Omega_{g x_0} = \Gamma_g \Omega_{x_0}$ for every $g \in G$ (and similarly for $\Omega_{g x_0}'$). Observe that for any $g_1,g_2$ such that $g_1 x_0 = g_2 x_0$ we have $\Omega_{g_1 x_0} =\Omega_{g_2 x_0}$ (and similarly for $\Omega_{g_1 x_0}' =\Omega_{g_2 x_0}'$). This defines the maps $\Omega: X \to V$ and $\Omega': X \to V$. We call $\Omega_X$  and $\Omega_X'$ the orbits of $\Omega_{x_0}$ and $\Omega_{x_0}'$ under $\Gamma_G$.
	
We call $\Lambda_\F \in V^*$ the set of all linear functionals giving values in $[0,1]$ on $\conv\left(\Omega_X\right)$ and $\Lambda_{\F'}' \in V^*$ the set of all linear functionals giving values in $[0,1]$ on $\conv\left(\Omega_X'\right)$. Necessarily $\sp(\Lambda_{\F'}) = \sp(\Lambda_{\F'}') = V^*$. $\Lambda_\F$ and $\Lambda_{\F'}'$ define two OPF sets $\F_X$ and $\F_X'$ respectively. 

We have constructed two systems $(X,G,\F_X)$ and $(X,G,\F_X')$ with associated maps $(\Omega,\Gamma,\Lambda)$ and $(\Omega',\Gamma,\Lambda')$ respectively, associated to the same representation $\Gamma$. We now show that these two systems are tomographically inequivalent, i.e. the mixed states $\conv\left(\Omega_X\right)$ and $\conv\left(\Omega_X'\right)$ are not isomorphic. In other words there is no linear invertible transformation mapping $\conv\left(\Omega_X\right)$ to $\conv\left(\Omega_X'\right)$.

	Consider an invertible linear transformation such that $L \Omega_X = \Omega_X'$ (if such a transformation exists, then it is well defined on the convex hulls and maps $\conv\left(\Omega_X\right)$ to $\conv\left(\Omega_X'\right)$). This implies that $L \Omega_x = \Omega_{gx}'$ for some fixed $g \in G$ and for all pure states $x \in X$. Let us redefine $L \mapsto \Gamma_{g^{-1}}$ such that the equality  $L \Omega_x = \Omega_{x}'$ for all $x \in X$ holds.

	Since the state spaces are transitive we have $\Omega_{x} = \Omega_{g x_0} = \Gamma_g \Omega_{x_0}$ for reference state $x_0$, and $\Omega_{x}' = \Omega_{g x_0}' = \Gamma_g \Omega_{x_0}'$. Using  $L \Omega_x = \Omega_x' \ \forall x \in X $ we obtain:
	\begin{align}
	L \Gamma_g \Omega_{x_0} = \Gamma_g \Omega_{x_0}' , \forall g \in G \ .
	\end{align}
	Moreover from $L \Omega_{x_0} = \Omega_{x_0}'$ we obtain 
	\begin{align}
	L \Gamma_g \Omega_{x_0} = \Gamma_g L \Omega_{x_0} , \forall g \in G \ .
	\end{align}
	Hence there is a transformation $L$ which commutes with the group action such that $L \Omega_{x_0} = \Omega_{x_0}'$. This is in contradiction with the assumption that the two reference states were not related by such a transformation. 
\end{enumerate}

\end{proof}

\section{Deformation of probabilistic structure}

\subsection{Proof of Theorem~\ref{thm:deformation}}\label{app:def_proof}

\begin{proof}
	Let $\Gamma^i: G \to {\rm GL}(V^i)$ be the representation of $G$ associated to $\F^i$, let $\Omega^i : X \to V^i$ be the representation of pure states, and let $\Lambda^i : \F^i \to (V^i)^*$ be the representation of OPFs, for $i=0,1$. 
	We can decompose the group action into (real) irreducible representations as $\Gamma^i = \bigoplus_j \Gamma^i_j$, $V^i = \bigoplus_j V^i_j$, $\Gamma^i_j : G \to {\rm GL} (V^i_j)$, where $j=0, 1, \ldots$
	Recall that there must be one (and only one) trivial irrep, which we label by $j=0$.
	Also, we can decompose the representation of pure states $\Omega^i = \bigoplus_j \Omega^i_j$ and $\Omega^i_j : X \to V^i_j$, and the representation of OPFs $\Lambda^i = \bigoplus_j \Lambda^i_j$ and $\Lambda^i_j : \F_j \to (V^i_j)^*$.
	Within this proof we follow the convention that 
	\begin{equation}\label{eq: normalisation}
	\langle \Omega^i_j(x), \Omega^i_j(x)\rangle_j = 1\ ,  
	\end{equation}
	for all $i,j,x$, where $\langle \cdot, \cdot \rangle_j$ is a group-invariant scalar product.
	This scalar product provides an isomorphism $(V^i_j)^* \cong V^i_j$.
	Using this decomposition we can write any $\f^i \in \F^i$ as 
	\begin{equation}\label{eq: effect}
	\f^i(x) = c^i_0(\f^i) +\sum_{j\geq 1} \langle \Lambda^i_j(\f^i), \Omega_j^i(x) \rangle_j\ ,  
	\end{equation}
	where we define 
\begin{equation}\label{def:c}
  c_j^i(\f^i) = \langle \Lambda^i_j (\f^i), \Lambda^i_j(\f^i) \rangle_j ^{1/2}\ .  
\end{equation}
	Schur's Lemma tells us that any non-trivial irreducible representation $\Gamma: G \to {\rm GL}(V)$ satisfies $\int_G dg\, \Gamma(g) = 0$.
Therefore, only the trivial rep survives the following average	
	\begin{align}\label{eq: f}
	\int_G dg\, \f^i(g x) 
	= \int_G dg \left[c^i_0(\f^i) + \sum_{j\geq 1} \langle \Lambda^i_j(\f^i), \Gamma^i_j(g)\, \Omega_j^i(x) \rangle_j \right]
	= c^i_0(\f^i) \ .
	\end{align}
Next we use another consequence of Schur's Lemma, namely that for any pair of (real) inequivalent irreducible representations $\Gamma: G \to {\rm GL}(V)$ and $\Gamma': G \to {\rm GL}(V')$ we have
\begin{align}\label{eq:laberl}
  \int_G dg\, 
  \langle \Lambda ,\Gamma(g) \Omega\rangle\, 
  \langle \Lambda' ,\Gamma'(g) \Omega'\rangle\,
  = 0\ ,
\end{align}
for any vectors $\Lambda, \Omega$ in $V$ and any vectors $\Lambda', \Omega'$ in $V'$.
Also, if the real irreducible representation $\Gamma: G \to {\rm GL}(V)$ has complexification with real structure then
\begin{align}
  \int_G dg\, 
  \langle \Lambda ,\Gamma(g) \Omega\rangle^2 
  &= 
  \int_G dg\, 
  \langle \Lambda ,\Gamma(g) \Omega\rangle\, 
  \langle \Omega ,\Gamma(g)^\dagger \Lambda\rangle
  \\ &= 
  \int_G dg\, 
  \langle \Lambda ,\Gamma(g) \Omega\rangle\, 
  \langle \Omega ,\Gamma(g^{-1}) \Lambda\rangle
  \\ &= 
  \frac 1 d 
  \langle \Lambda ,\Lambda\rangle\, 
  \langle \Omega ,\Omega\rangle\ , 
\end{align}
where $d$ is the dimension of $V$ and $\Lambda, \Omega$ are any vectors in $V$.

The case where the real irreducible representation $\Gamma: G \to {\rm GL}(V)$ acts on the complexified space $\mathbb C V \cong W \oplus \bar W$ as the direct sum of two complex irreducible representations $\gamma \oplus \bar\gamma$, with complex structure and one being the dual of the other.
For what follows we are interested in elements of $W \oplus \bar W$ of the form $\Omega = \omega \oplus \bar\omega$ where $\bar\omega$ is the complex conjugate of $\omega$, and analogously for $\Lambda = \lambda \oplus \bar\lambda$.
Proceeding as above and using Schur's Lemma we obtain
\begin{align}
  \int_G dg\, 
  \langle \Lambda ,\Gamma(g) \Omega\rangle^2 
  &= 
  \int_G dg\, 
  \langle \Lambda ,\Gamma(g) \Omega\rangle\, 
  \langle \Omega ,\Gamma(g)^\dagger \Lambda\rangle
  \\ &= 
  \int_G dg\, 
  \langle \Lambda ,\Gamma(g) \Omega\rangle\, 
  \langle \Omega ,\Gamma(g^{-1}) \Lambda\rangle
  \\ \nonumber &= 
  \int_G dg\, 
  \left(\langle \lambda ,\gamma(g) \omega\rangle +\langle \bar\lambda ,\bar\gamma(g) \bar\omega \rangle\right)\, 
  \left(\langle \omega ,\gamma(g^{-1}) \lambda\rangle + \langle \bar\omega ,\bar\gamma(g^{-1}) \bar\lambda\rangle\right)
  \\ &= 
  \frac 2 {d}\langle \lambda ,\lambda\rangle \langle \omega ,\omega\rangle 
  +\frac 2 {d} \langle \bar\lambda ,\bar\lambda\rangle \langle \bar\omega ,\bar\omega\rangle 
\end{align}
where $d/2$ is the dimension of $W$ (and $\bar W$).
The structure of the vector $\Omega$ implies
\begin{equation}
  \langle\Omega,\Omega\rangle
  =
  \langle\omega,\omega\rangle +\langle\bar\omega,\bar\omega\rangle 
  = 2 \langle\omega,\omega\rangle
  = 2 \langle\bar\omega,\bar\omega\rangle \ ,
\end{equation}
and analogously for $\Lambda$. 
Which in turn implies
\begin{align}\label{eq:complex g}
  \int_G dg\, 
  \langle \Lambda ,\Gamma(g) \Omega\rangle^2 
  = 
  \frac 1 {d}\langle \Lambda ,\Lambda\rangle \langle \Omega ,\Omega\rangle \ .
\end{align}
In summary, the above average is independent of whether the complexification of representation $\Gamma$ is reducible or not.

The remaining case, where $W$ has quaternionic structure, need not be considered, because Lemma~7 proves that it never appears in Gelfand systems.
Next we use the identities \eqref{eq:laberl} and \eqref{eq:complex g}, the normalisation convention \eqref{eq: normalisation} and definition \eqref{def:c} to calculate the following average
\begin{align}
    \nonumber
	\int_G dg\, [\f^i(g x)]^2 
	=\ &c^i_0(\f^i)^2 
	+\sum_{j\geq 1} \frac {1} {d^i_j}\, 
	\langle \Lambda^i_j(\f^i), \Lambda^i_j(\f^i) \rangle \,
	\langle \Omega^i_j(x), \Omega^i_j(x)\rangle_j	
	\\ \label{eq: f2} 
	=\ 
	\sum_{j} \frac {c^i_j(\f^i)^2} {d^i_j}	\ ,
\end{align}
	where $d^i_j = {\rm dim} V^i_j$.

	Now, suppose that $\F^0$ and $\F^1$ are unrestricted and tomographically-inequivalent probabilistic structures of $(G,X)$.
	Then, without loss of generality, we assume that irrep $\Gamma^0_1$ is inequivalent to irrep $\Gamma^1_{j}$ for all $j$.
	Also, let us choose $\f^0 \in \F^0$ so that its corresponding vector $\Lambda^0(\f^0)$ only has support on the subspaces $j=0,1$, that is
\begin{equation}\label{eq:effect2}
	\f^0(x) = 
	c^0_0(\f^i) + \langle \Lambda^0_1(\f^0), \Omega_1^0(x) \rangle_1 
	\ .
\end{equation}
This choice is possible because the probabilistic structure $\F^0$ is unrestricted.
Also, due to the fact that we are obtaining a lower bound for the distance $\mathcal D(\F^0,\F^1)$, we do not need to maximise over OPFs $\f^0\in\F^0$, it is enough to pick any one.
	Next, let us lower-bound the distance between the chosen OPF $\f^0 \in \F^0$ and any given OPF $\f^1 \in \F^1$ as
	\begin{align}\nonumber
	\dist(\f^0,\f^1) 
	&= \max_{x \in X} |\f^0(x) -\f^1(x)|
	\geq \int_G dg\, \left|\f^0(g x_0) -\f^1(g x_0) \right| \\
	&\geq \int_G dg\, \left[\f^0(g x_0) -\f^1(g x_0) \right]^2 \ ,
	\end{align}  
	for any $x_0 \in X$.
	This bound is independent of whether the probabilistic structures $\F^0$ and $\F^1$ are equivalent or not.
	Now, using Schur's lemma as in \eqref{eq: f} and \eqref{eq: f2} we obtain
\begin{align}\nonumber
	\dist(\f^0,\f^1) 
	&\geq \sum_{j=0}^1 \frac {c^0_j(\f^0)^2} {d^0_j}
	+ \sum_{j} \frac {c^1_j(\f^1)^2} {d^1_j}
	- 2\, c^0_0(\f^0)\, c^1_0(\f^1)
\\ \nonumber
	&= \frac {c^0_1(\f^0)^2} {d^0_1}
	+ \sum_{j\geq 1} \frac {c^1_j(\f^1)^2} {d^1_j}
	+ \left[ c^0_0(\f^0) - c^1_0(\f^1) \right]^2
\\
	&\geq  \frac {c^0_1(\f^0)^2} {d^0_1}
	+ \sum_{j\geq 1} \frac {c^1_j(\f^1)^2} {d^1_j} \ ,
\end{align}  
	for any $\f^1 \in \F^1$, therefore
\begin{align}\label{eq:bound}
    \mathcal D(\F^0,\F^1) 
    \geq 
	\min_{\f^1 \in \F^1} \dist(\f^0,\f^1) 
	\geq  \frac {c^0_1(\f^0)^2} {d^0_1}\ .
\end{align}  
Again, using the fact that we can pick any $\f^0\in\F^0$, we further restrict \eqref{eq:effect2} to 
\begin{equation}\label{eq:effect3}
	\f^0(x) = 
	\frac 1 2 + \frac 1 2 \langle \Omega_1(x_0), \Omega_1(x) \rangle 
	\in[0,1]\ ,
\end{equation}
for any fixed state $x_0\in X$.
The above $\f^0$ is an OPF because the normalisation condition \eqref{eq: normalisation} implies that it takes values within $[0,1]$.
This chosen OPF \eqref{eq:effect3} has $c^0_1(\f^0) = 1/2$, which when substituted in \eqref{eq:bound} gives
\begin{align}
	\mathcal D(\F^0,\F^1)
	\geq  
	\frac {1} {4 d^0_1}\ ,
\end{align}  
which is the statement of Theorem~\ref{thm:deformation}.
\end{proof}

\subsection{Proof of Theorem~\ref{thm:deformation2}}\label{app:def_prob_struc}

\subsubsection{Proof of  Theorem~\ref{thm:deformation2} $i.$}

\begin{proof}
Let $\F^0$ of dimension $\dim \R[\F_0] = d_0$ be an unrestricted probabilistic structure of $(G,H)$ with associated representation $\Gamma_G^0$ such that all $H$-invariant vectors are related by invertible transformations which commute with $\Gamma_G^0$. 

Consider two such vectors $v_H$ and $v_H'$ where $v_H' = Lv_H$, with $L \in \GL(\R[\F_0])$. We can generate state spaces with extremal points $G/H$ by applying $\Gamma_G$ to the reference states:
\begin{align}
&\Omega_{G/H} = \Gamma_G v_H \ ,\\
&\Omega_{G/H}' = \Gamma_G v_H'  = \Gamma_G L v_H = L \Omega_{G/H} \ .
\end{align}
Both the state spaces $\conv\left(\Omega_{G/H} \right)$ and $\conv\left(\Omega_{G/H}' \right)$ are related by an invertible transformation and hence are equivalent as convex sets. In other words they correspond to tomographically equivalent probabilistic structures. Since we are only considering unrestricted probabilistic structures they both generate the same state space corresponding to $\F_0$ (the linear transformation $L$ is just a change of co-ordinates, including rescalings). For probabilistic structures whose $H$-invariant vectors are  all related to each other by invertible transformations which commute with $\Gamma_G$ there is a unique unrestricted probabilistic structure associated to $\Gamma_G$.

Therefore the only other probabilistic structures $\F_1$ such that $\dim \R[\F_1] = d_0$ (if they exist) are associated to different representations $\Gamma^1_G$. The largest dimensional irreducible representation they can differ by is of dimension $d_0-1$ (corresponding to the case where either $\Gamma^0$ or $\Gamma^1$ consists of a $d_0-1$ dimensional irreducible representation and the trivial and the other representation consists of the trivial representation and an inequivalent $d_0 - 1$ irreducible representation or some reducible representation of total dimension $d_0 - 1$). Therefore by Theorem~\ref{thm:deformation} the distance between the two probabilistic structures is lower bounded by $\frac{1}{4(d_0-1)}$. Any other possibility (i.e. both consist of a trivial representation and different reducible ones would give a lower bound which is higher).
\end{proof}

\subsubsection{Proof of  Theorem~\ref{thm:deformation2} $ii.$}

Let $\F^0$ of dimension $\dim \R[\F_0] = d_0$ be an unrestricted probabilistic structure of $(G,H)$ with associated representation $\Gamma_G^0$ such that there are pairs of $H$-invariant vectors not related by invertible transformations which commute with $\Gamma_G^0$. These $H$-invariant vectors belong to an $H$-invariant subspace $V_H$.

Take $\Omega_H \in V_H$ and $\Omega_H^t = L^t \Omega_H$ where $L^t = e^{iRt}$ with $R \in \mathfrak{gl}\left(V_H \right)$. Take $\Omega_H^\epsilon$ for small $\epsilon > 0$ . We then expand:
\begin{align}
\Omega_H^\epsilon = (\I + \epsilon R + ... ) \Omega_H =  \Omega_H + \epsilon R \Omega_H + O(\epsilon^2) \ .
\end{align}
We consider a first order approximation and ignore terms of order $\epsilon^2$ and higher.

The image of a point $x = gH$ is $\Omega_{gH} = \Gamma_g \Omega_H$ and  $\Omega_{gH}^\epsilon = \Gamma_g \Omega_H^\epsilon$:
\begin{align}
\Omega_{gH}^\epsilon = \Omega_{gH} + \epsilon \Gamma_g R \Omega_H  \ .
\end{align}
Let us take an OPF $\f^0 \in \F_0$ with associated effect $\Lambda_{\f^0}$ and an OPF $\f^1 \in \F_\epsilon$  with associated effect  $\Lambda_{\f^1}^\epsilon$:
\begin{align*}
\f^0(gH) = \Lambda_{\f^0} \cdot \Omega_{gH} \ , \\
\f^1(gH) = \Lambda_{\f^1}^\epsilon \cdot \Omega_{gH}^\epsilon \ .
\end{align*} 
\begin{align}
\dist(\f^0,\f^1) & = \max_{x \in X} |\f^0(x) -\f^1(x)| = \max_{x \in X} | \Lambda_{\f^0}\cdot \Omega_x - \Lambda_{\f^1}^\epsilon \cdot \Omega_x^\epsilon| \\ 
& =  \max_{g \in G} |(\Lambda_{\f^0} - \Lambda_{\f^1}^\epsilon ) \cdot  \Omega_{gH} -  \epsilon  \Lambda_{\f^1}^\epsilon \cdot  (\Gamma_g R \Omega_H) | \ .
\end{align}
For a fixed $\f_0$ we want to find an $\f^1$ which minimizes the distance.
\begin{align}\label{eq:dist_f0_f1}
\min_{\f^1 \in \F^\epsilon} \dist(\f^0,\f^1)  = \min_{\f^1 \in \F^\epsilon}   \max_{g \in G} |(\Lambda_{\f^0} - \Lambda_{\f^1}^\epsilon ) \cdot \Omega_{gH} -  \epsilon  \Lambda_{\f^1}^\epsilon \cdot ( \Gamma_g R \Omega_H) | \ .
\end{align}
We observe that that the transformations $\Gamma_g$ and $R$ leave the normalisation degree of freedom (component in the normalisation subspace $V_0$) unchanged, therefore if we take $\f^0 = \u^0$ then we can choose $\f^1 = \u^1$ to obtain $\dist(\u^0,\u^1) = 0$.

For any effect with support outside the normalisation subspace $V_0$ the expression~\eqref{eq:dist_f0_f1} is not linear in $\Omega_{gH}$ (due to the last term) and as such there is no choice of $\Lambda_{\f^1}$ which will make $(\Lambda_{\f^0} - \Lambda_{\f^1}^\epsilon )$ cancel out with $\epsilon  \Lambda_{\f^1}^\epsilon  \Gamma_g R \Omega_H $ for all $g \in G$, which is needed to make the distance go to 0. Hence for $\f^0$ not proportional to $\u^0$ we have the bound:
\begin{align*}
\min_{\f^1 \in \F^\epsilon} \dist(\f^0,\f^1) > 0 \ .
\end{align*}
We observe that every $\Omega^\epsilon_x$ is $\epsilon$ close to $\Omega_x$, therefore any $\Lambda_{\f^0}$ which is not $\epsilon-$tight on $\Omega_X$ (i.e. which gives values in $[0+ \epsilon, 1 - \epsilon]$) will be valid on $\Omega^\epsilon_X$. 
Let us minimize the distance $\min_{\f^1 \in \F^\epsilon} \dist(\f^0,\f^1)$ for the choice of $\f^0$ not $\epsilon-$tight. We observe that by choosing $\Lambda_{\f^1}^\epsilon = \Lambda_{\f^0}$ we obtain:
\begin{align}
\max_{g \in G} |(\Lambda_{\f^0} - \Lambda_{\f^1}^\epsilon ) \cdot  \Omega_{gH} -  \epsilon  \Lambda_{\f^1}^\epsilon \cdot  (\Gamma_g R v_H) | = |\epsilon  \Lambda_{\f^1}^\epsilon  \cdot (\Gamma_g R \Omega_H) | \leq  \epsilon \ ,
\end{align}
which implies that 
\begin{align}\label{eq:dist_epsilon}
\min_{\f^1 \in \F^\epsilon} \dist(\f^0,\f^1) \leq \epsilon \ .
\end{align}

Finally if $\f^0$ is $\epsilon-$tight (i.e. has some values in $[0, \epsilon]$ or $[1-\epsilon, 1]$), one can take $\Lambda_{\f^0}$ and add $\epsilon$ noise (i.e. take $\Lambda_{\f^0}' = (1- \epsilon) \Lambda_{\f^0} + \epsilon \Lambda_{\u^0}$) in order to obtain an effect $\Lambda_{\f^0}'$ which is valid on $\Omega^\epsilon_X$:

\begin{align}
& \max_{g \in G} |(\Lambda_{\f^0} - \Lambda_{\f^0}') \cdot  \Omega_{gH} -  \epsilon  \Lambda_{\f^0}' \cdot   (\Gamma_g R \Omega_H) | \\
&  = \min_{\f^1 \in \F^\epsilon}   \max_{g \in G}  |\epsilon - \epsilon  \Lambda_{\f^1}^\epsilon \cdot  (\Gamma_g R \Omega_H )| \leq 2 \epsilon \ .
\end{align}
Therefore 
\begin{align}
\min_{\f^1 \in \F^\epsilon} \dist(\f^0,\f^1) \leq 2 \epsilon \ .
\end{align}
This, together with Equation~\eqref{eq:dist_epsilon}, implies that
\begin{align}
\max_{\f^0 \in \F^0} \min_{\f^1 \in \F^\epsilon} \dist(\f^0,\f^1) \leq 2 \epsilon\ .
\end{align}

\subsection{Proof of Lemma~\ref{lem:deformation3}}\label{app:deformation3}

\begin{proof}

Two convex sets $S_0$ and $S_1$ with pure states $X$ linearly embedded in isomorphic spaces $V_0 \cong V_1 \cong \R^d$ (i.e. for which there exist injective maps $\Omega^0: X \to \R^d$ and $\Omega^1: X \to \R^d$ and $S_i = \conv(\Omega^i(X))$) can be continuously deformed one into the other if there exists a \emph{path connected map} $M_{0 \to t}: \Omega_X^0 \to \Omega^t_X$  in the space $\fF_d$ (of $\Omega$ maps with codomain $\R^d$) parametrised by $t \in [0,1]$.  $M_{0 \to t} \left( \Omega^1_X \right) = \Omega_X^t$ is an embedding of $X$ in $\R^n$ for every $t \in [0,1]$. We observe that $M_{0 \to t}$ is defined just on the extremal points of $\Omega^0_X$ (and is invertible), but cannot be extended to $\conv\left(\Omega_X^0\right)$ by Lemma~\ref{lem:non-linear_def}. A connected path on $\fF_d$ (the space of $\Omega$ maps with codomain $\R^d$) is a continuous map from $[0,1] \subset \R$ (with the usual topology) to $\fF_d$ (with the topology inherited from the metric $D_\sym$).

We now show that for any non-rigid probabilistic structure one can find probabilistic structures which can be continuously deformed into one another, whilst rigid probabilistic structures cannot be continuously deformed.

In the following we make use of the isomorphism $X \cong G/H$ and label points in $X$ by their stabilizer subgroup. The point $x$ with stabilizer subgroup $H$ is written as $H$, and any point $x' = gx$ is written as $gH$.

A probabilistic structure $\F_X$ is non-rigid if the associated representation $\Gamma$ contains an irreducible spherical representation $\Gamma$ with inequivalent $H$-invariant vectors $v_H^0$ and $v_H^1$. Two $H$-invariant vectors $v_H^0$ and $v_H^1$ are inequivalent if they are not related by an invertible transformation which commutes with $\Gamma_G$.

We show that a non-rigid probabilistic structure $\F_X^0$ associated to an irreducible representation $\Gamma$ can be continuously deformed. The case of a non-rigid probabilistic structure $\F_X^0$ associate to a reducible representation follows since one can deform the full reducible probabilistic structure by deforming the subspace of $\F_X^0$ acted on by the irreducible representation with inequivalent $H$-invariant vectors.

The embedded pure state $\Omega_X^0$ associated to $\F_X^0$ can be generated as $\Omega_X^0 = \{\Gamma_g v_H^0| g \in G\}$, where $v_H^0$ is an $H$-invariant vector.  Let $v_H^1$ be an inequivalent $H$-invariant vector. Any vector in $\spann (v_H^0, v_H^1)$ is $H$-invariant, moreover any two vectors in $\spann (v_H^0, v_H^1)$ which are not proportional will generate inequivalent state spaces under $\Gamma_G$.   Let us call $\Omega_X^1 = \{\Gamma_g v_H^1| g \in G\}$ the embedded sets of pure states generated from the reference state $v_H^1$. We now show that $\Omega_X^0$ can be continuously deformed to $\Omega_X^1$.

Consider a rotation $R \in \GL (\spann (v_H^0, v_H^1))$ such that $Rv_H^0 = v_H^1$. 

%
Let us write  $R(t) = e^{i Kt}$ with $R(1) = R$ for some non-unique generator $K$.  

We now show the existence of a connected path in $\fF_d$ from $\Omega_X^0$ to $\Omega_X^1$, the two sets pure states generated by two non-equivalent $H$-invariant vectors $v_H^0$ and $v_H^1$.

Let us define $\Omega_X^t$ the state space generated by the $H$-invariant reference vector $v_H^t = R(t) v_H^0$.

We now define $\gamma: [0,1] \to \fF_d$, where $\gamma(t) = \Omega_X^t$.  We observe that $\gamma(0) = \Omega^0_X$ and $\gamma(1) = \Omega^1_X$. We now show that $\gamma$ is a continuous function (where $\fF_d$ has the topology induced by the metric $D_\sym$ of Equation~\eqref{eq:dist_prob})

$\gamma$ is continuous at a point $t_0 \in [0,1]$ if for every $\epsilon >0$ it is the case that if $\D_\sym(\Omega^{t_0}_X, \Omega^t_X)< \epsilon$ (for $t \in [0,1]$) then we can find a $\delta$ such that $|t - t_0|< \delta$ implies $\D_\sym(\Omega^{t_0}_X, \Omega^t_X)< \epsilon$. 

For $\epsilon$ large (but $\leq 1$) this holds since one can just take $\delta$ to be small, in which case $\D_\sym(\Omega^{t_0}_X, \Omega^{t_0}_X) \leq 2 \delta$ by the proof of Theorem~\ref{thm:deformation2} $ii.$ which is less than or equal to $\epsilon$.

If $\epsilon <<1$ (and greater than $0$) then setting $\delta = \frac{\epsilon}{2}$ gives $|t - t_0|<  \frac{\epsilon}{2}$ which implies  $\D_\sym\left(\Omega^{t_0}_X, \Omega^t_X\right) < \epsilon$ using the proof of Theorem~\ref{thm:deformation2} $ii.$.

This holds for all $t_0 \in [0,1]$ hence $\gamma$ is a continuous function from $[0,1]$ to $\fF_d$ with extremal points $\Omega^0_X$ and $\Omega^1_X$. Hence there is a connected path from $\Omega^0_X$ and $\Omega^1_X$ in the space $\fF_D$ of deformation maps, meaning that $\Omega_X^0$ can be continuously deformed to $\Omega_X^1$.

In the case where $\F_X^0$ is a rigid probabilistic structure it is at a finite bounded distance from every other probabilistic structure in $\fF_d$ by definition. Consider the associated set of embedded pure states $\Omega_X^0$ and take an arbitrary path $\gamma(t)$ to an arbitrary $\Omega_X^1 \in \fF_d$. $\gamma(t)$ is not continuous at $t = 0$, since for $\epsilon$ lower than the finite bounded distance from $\F_X^0$ to any other probabilistic structure, there is no $\delta$ such that $|t|< \delta$ implies $\D_\sym\left(\Omega^{0}_X, \Omega^t_X\right) < \epsilon$.

\end{proof}

\section{Proof of Lemma~\ref{lem:rep_thry_ex} part 2}\label{app:rep_thry_ex}

\begin{proof}

We make use of Proposition 26.24 of~\cite{Fulton_representation_1991} which we translate slightly:

\begin{lemma}
	Any irreducible representation of $\SU(d)$ with Dynkin indices $j = (j_1,..., j_{d-1})$ is complex if $j_i \neq j_d-i$ for any $i$, real if $j_i = j_{d-i}$ for all $i$ and $n$ is odd, or $n = 4k$, or $n = 4k+2$ and $j_{2k +1}$ is even, and quaternionic if  $j_i = j_{d-i}$ for all $i$ and $j_{2k +1}$ is odd.
\end{lemma}

We translate the partitions \eqref{eq:rep_one} and \eqref{eq:rep_two} into Dynkin indices: 

When $m =n$:
\begin{equation}
j = ( b _1 - b_2 ,  b_2 -b_3 , ... , 2 b_m, b_{m-1} - b_m , ... , b_1-b_2)
\end{equation}
where this has $m+n-1$ entries.

For all $i$ $j_i = j_{m+n-i}$, hence for $m+n$ odd and $m+n = 4 k$ these are all real. For $m +n = 4k +2$ we observe that $j_{2 k +1}$ (the central entry) is $2 b_m$ and hence even. This implies  the representation is real.

When $n \geq m + 1 $:
\begin{equation}
\lambda = (b _1 - b_2 , b_2 -b_3 , ... ,  b_m, \underbrace{0, .... , 0}_{{\rm times } \  n - m -1},  b_m , ...,b_2-b_3 , b_1-b_2)
\end{equation}
where this has $m+n-1$ entries. 

For all $i$ $j_i = j_{m+n-i}$, hence for $m+n$ odd and $m+n = 4 k$ these are all real. For $m +n = 4k +2$ we observe that $j_{2 k +1}$ (the central entry) is $0$ and hence even. This implies  the representation is real.

\end{proof}

\end{document}